\documentclass[preprint2]{aastex}

\received{} \accepted{}

\shorttitle{Molecular Hydrogen kinematics in Cepheus A}
\shortauthors{Hiriart, Salas \& Cruz-Gonz\'alez}

\begin{document}

\title{Molecular Hydrogen Kinematics in  Cepheus A\altaffilmark{1}}

\author{David Hiriart\altaffilmark{2} and  Luis Salas\altaffilmark{2} }

\affil{Instituto de Astronom\'{\i}a (Unidad Ensenada), Universidad
Nacional Aut\'onoma de M\'exico}
\affil{Apdo. Postal 877, 22830 Ensenada, B.C. M\'exico}
\email{hiriart@astrosen.unam.mx, salas@astrosen.unam.mx}
\and
\author{ Irene Cruz-Gonz\'alez\altaffilmark{2}}
\affil{Instituto de Astronom\'{\i}a, UNAM}
\affil{Circuito Exterior C.U., 04510 M\'exico, D. F., M\'exico}
\email{irene@astroscu.unam.mx}

\altaffiltext{1}{\protect\raggedright Based on observations made at
the 2.1 m telescope of the Observatorio Astron\'omico Nacional at San
Pedro M\'artir, B.C., M\'exico.}

\altaffiltext{2}{Observatorio Astron\'omico Nacional, San Pedro
M\'artir, B.C., M\'exico.}

\begin{abstract}
  We present the radial velocity structure of the molecular hydrogen
outflows associated to the star forming region Cepheus A. This
structure is derived from the doppler shift of the H$_2$ $v$=1--0
$S$(1) emission line obtained by Fabry--P\'erot spectroscopy. The East
and West regions of emission, called Cep~A~(E) and Cep~A~(W), show
radial velocities in the range of -20 to 0~km~s$^{-1}$ with respect to
the molecular cloud.  Cep~A~(W) shows an increasing velocity with
position offset from the core indicating the existence of a possible
accelerating mechanism.  Cep~A~(E) has an almost constant mean radial
velocity of -18~km~s$^{-1}$ along the region although with a large
dispersion in velocity, indicating the possibility of a turbulent
outflow.  A detailed analysis of the Cep~A~(E) region shows evidence
for the presence of a Mach disk on that outflow. Also, we argue that
the presence of a velocity gradient in Cep~A~(W) is indicative of a
C-shock in this region. Following \citet{Riera03}, we analyzed the
data using wavelet analysis to study the line width and central radial
velocity distributions. We found that both outflows have complex
spatial and velocity structure characteristic of a turbulent flow.
\end{abstract}


\maketitle

\section{Introduction}
\label{sec:Introduction}

  Cepheus A is the densest core within the Cepheus OB3 molecular cloud
complex \citep{Sargent77} and a massive star forming region. It
contains a deeply embedded infrared source which generates a total
luminosity of $\sim2.4\times 10^4 L_\odot$ \citep{Koppenaal79}.

  Two main regions of ionized and molecular gas about 2 arc minutes
apart and oriented roughly in the east-west direction, have been
detected, Cepheus A East \citep{Bally82} and Cepheus A West (Simon \&
Joyce 1983; Garay et al. 1996). The first molecular hydrogen map of
both regions using Fabry-P\'erot spectroscopy was presented by
\citet{Doyon88}.  Later, \citet{Hartigan86} obtained high resolution
images in the $v$=1--0 $S$(1) line emission of H$_2$ from Cepheus A
West. The two Cepheus A regions with molecular hydrogen emission show
quite different compositions.

  The eastern region (hereon Cep~A~(E)) hosts one of the first
detected CO bipolar molecular outflows \citep{Rodriguez80}. High
resolution observations show a more complex outflow of a quadrupole
nature \citep{Torrelles93}.  \citet{Torrelles93} suggested that the
source Cep A East:HW2 \citep{Hughes84} is powering the PA=45$\degr$
outflow but it is not clear if the powering source of the
PA=115$\degr$ outflow is Cep A East:HW3 or another source
(L. F. Rodr\'{\i}guez, personal communication). Observations of
$^{12}$CO, CS, and CSO give evidence of multiple episodes of outflow
activity \citep{Narayanan96}. Observations by \citet{Codella03} of
H$_2$S and SO$_2$ confirm the presence of multiple outflows. Highly
variable H$_2$O and OH masers, commonly associated to young
stellar objects, are surrounded by very dense NH$_3$ condensations
that probably redirect the outflow into a quadrupole structure
\citep{Torrelles93, Narayanan96}.

   The western region (hereon Cep~A~(W)) contains several radio
continuum sources at 3~cm \citep{Garay96} and \citet{Hartigan86}
identified a region of several Herbig-Haro objects known as HH 168
(GGD~37) with large radial velocities and line widths. A bipolar
outflow of CO with overlapping red- and blue-shifted lobes is
associated to this region \citep{Bally82,Narayanan96}. It should be
pointed out that the energy source of Cep~A~(W) remains elusive
\citep{Raines00, Garay96, Torrelles93, Hartigan85}.

   Although the two regions Cep~A~(E) and Cep~A~(W) could constitute a
single large structure outflow, several authors have presented
evidence which suggests that Cep~A~(W) may be an independent region of
activity, distinct from Cep~A~(E) \citep{Raines00,Garay96,Hartigan85}.

   In this paper we present the radial velocity structure of the
Cepheus~A molecular hydrogen outflows obtained from the H$_2$ $v$=1--0
$S$(1) doppler shifted emission line at 2.122~\micron~ measured by
scanning Fabry-P\'erot spectroscopy.  Due to the complexity found in
the velocity structures, we decided to study the kinematics by using
an asymmetric wavelet analysis following \citet{Riera03}, who used
this method to study H$\alpha$ Fabry-P\'erot observations of the
HH~100 jet.

  Our results show that the two regions represent turbulent H$_2$
outflows with significant differences from a kinematic point of
view. A detailed analysis of the Cep A (E) region provides evidence
for the presence of a Mach disk near the tip of the outflow.

 In \S~\ref{sec:Observations} we described the observations.  From
this data, in \S~\ref{sec:Results} we generate a doppler shift H$_2$
image, radial velocity, velocity gradient and line width maps, and
study the flux-velocity diagrams \citep{Salas02}. By using the
asymmetric wavelet transform, the clumpy structures of both regions of
Cepheus~A are kinetically analyzed and discussed
in~\S~\ref{sec:Wavelet}. The conclusions are then summarized and
presented in~\S~\ref{sec:Conclusions}.

\section{Observations}
\label{sec:Observations}

   In October 5, 1998, we observed the Cepheus~A region with the 2.1~m
telescope of the Observatorio Astron\'omico Nacional at San Pedro
M\'artir,~B.C. in Mexico.

   The measurements were obtained with the CAMILA near-infrared
camera/spectrograph \citep{Cruz-Gonzalez94} with the addition of a
cooled tunable Fabry-P\'erot interferometer (located in the collimated
beam of the cooled optical bench), and a 2.12~\micron~ interference
filter. A detailed description for the infrared scanning
Fabry-P\'erot instrumental setup is presented by \citet{Salas99}.

  The Fabry-P\'erot has a spectral resolution of 24 km~s$^{-1}$ and
to restrict the spectral range for the $v$=1--0 $S$(1) H$_2$ line
emission, an interference filter (2.122~\micron~ with $\Delta \lambda
$=~0.02~\micron~) was used.  The bandwidth of this filter allows 11
orders of interference. Only one of these orders contain the
2.122~\micron~line and the remaining orders contribute to the observed
continuum.  The spatial resolution of the instrumental array is $\sim$
0.86\arcsec~ pixel$^{-1}$.  The field of view allows to cover a
3.67\arcmin~ $\times$ 3.67\arcmin~ region, which corresponds to 0.7
$\times$ 0.7 pc$^2$ at the adopted distance of 725 pc
\citep{Johnson57}. With this field of view, one set of images was
required for the eastern portion and another for the western region of
Cepheus A.

  Images of each region of interest were obtained at 26 etalon
positions, corresponding to increments of 9.82 km~s$^{-1}$.  The
observing sequence consists of tunning the etalon to a new position
and imaging the source followed by a sky exposure at an offset of
5\arcmin~ south from the source. The integration time of 60~s per
frame was short enough to cancel the atmospheric lines variations at
each etalon position, but long enough to obtain a good
signal-to-noise ratio.  Images were taken under photometric
conditions with a FWHM of 1.6\arcsec.

  Spectral calibration was obtained by observing the line at 2.1332885
\micron~ of the Argon lamp at each position of the etalon, giving a
velocity uncertainty of 1~km~s$^{-1}$ in the wavelength fit. A set of
high- and low-illumination sky flats were obtained for
flat-fielding purposes.

  We reduced the data to obtain the velocity channel images using the
software and the data reduction technique described in
\citet{Salas99}.

\section{Results}
\label{sec:Results}

\subsection{H$_2$ Velocity Maps}
\label{H_2_Velocity Maps}

\begin{figure*}
\begin{center}
\includegraphics[width=\textwidth]{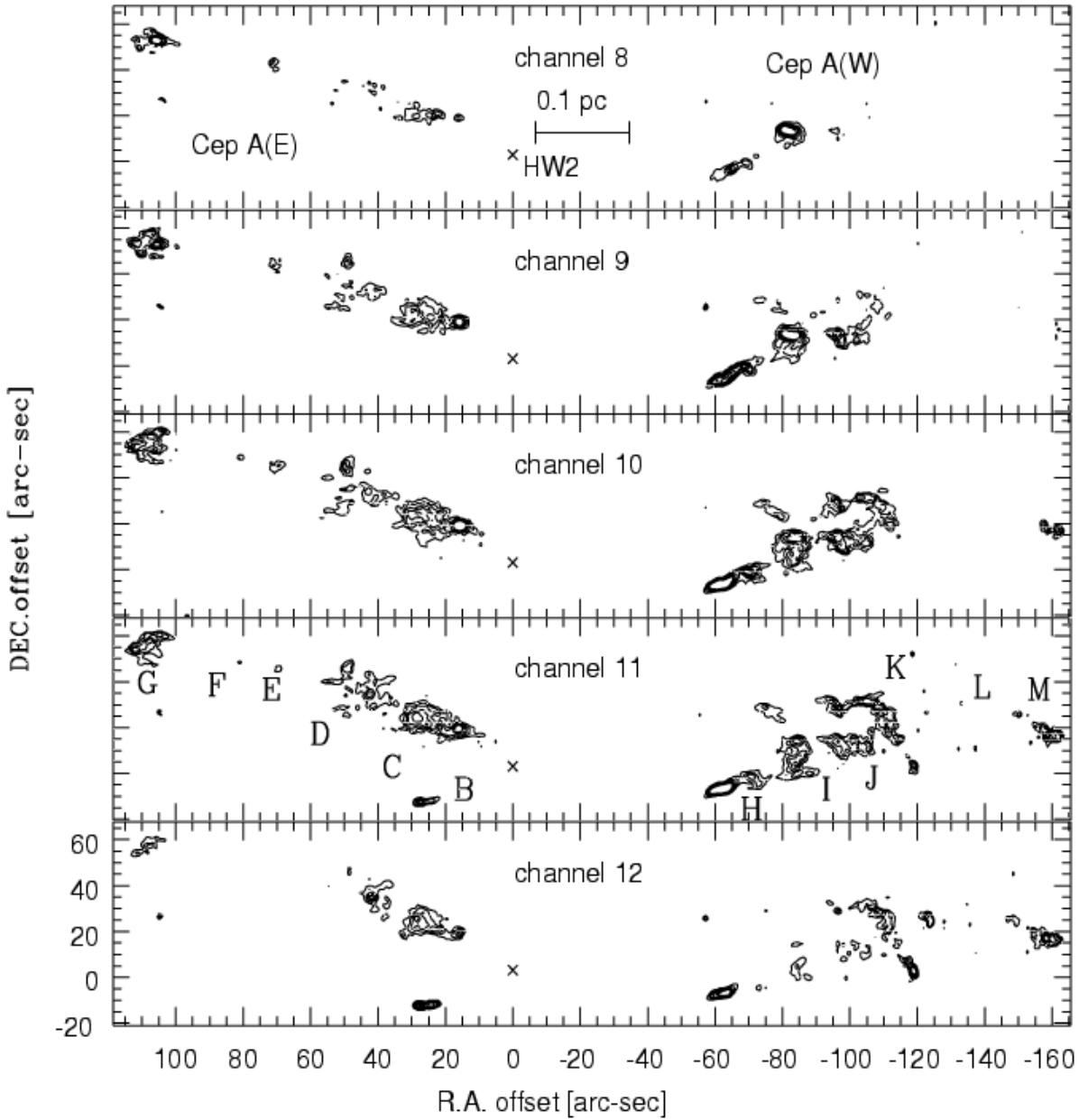}
\end {center}
\caption{ Velocity channel maps of the H$_2$ $v$=1--0 $S$(1) line
emission in Cepheus A, showing the radial velocity structure of the
two regions Cep~A~(E) and Cep~A~(W).  The velocity varies from -42.1
km~s$^{-1}$ in channel~8 ({\it top}) to -2.9 km~s$^{-1}$ in channel~12
({\it bottom}), the channel width is 9.8 km~s$^{-1}$.  The scale of
the region is shown at the top panel. The contour intervals indicate
the emission intensity in the range of 0.54 to 1.5~counts/sec and have
the same increment values of 0.2~counts/sec. The cross indicates the
position of the 6~cm peak of HW2 at R.A.~=~22h45m17.9s and
DEC.~=~+62\degr~01\arcmin~49\arcsec~ (2000), see
\citet{Hughes84}. Offsets are referred to this reference position.
\label{FIG:Channel-map}}
\end{figure*}

  Velocity channel images were individually obtained for Cep~A~(E) and
Cep~A ~(W) from the position-velocity cube data. For each pixel on
the image we subtracted a continuum intensity level calculated from
the median of the channels with no H$_2$ emission.  H$_2$ $v$=1--0
$S$(1) line emission was detected in velocity channels -40 to
0~km~s$^{-1}$ in Cep~A~(E) and in channels -40 to 10 km~s$^{-1}$ in
Cep~A~(W). The two sets of maps were pasted together to create
velocity channel maps of the complete
region. Figure~\ref{FIG:Channel-map} shows five of these maps (8 to
12) covering Local Standard of Rest (LSR) velocities from
-42~km~s$^{-1}$ to -3~km~s$^{-1}$. Cep~A~(E) shows H$_2$ emission in
six separated clumps of emission (labeled B--G in
Fig.~\ref{FIG:Channel-map}), while Cep~A~(W) the emission may be
distinguished in six regions (labeled H--M).

%figura 1

\begin{figure*}
\begin{center}
\includegraphics[width=\textwidth]{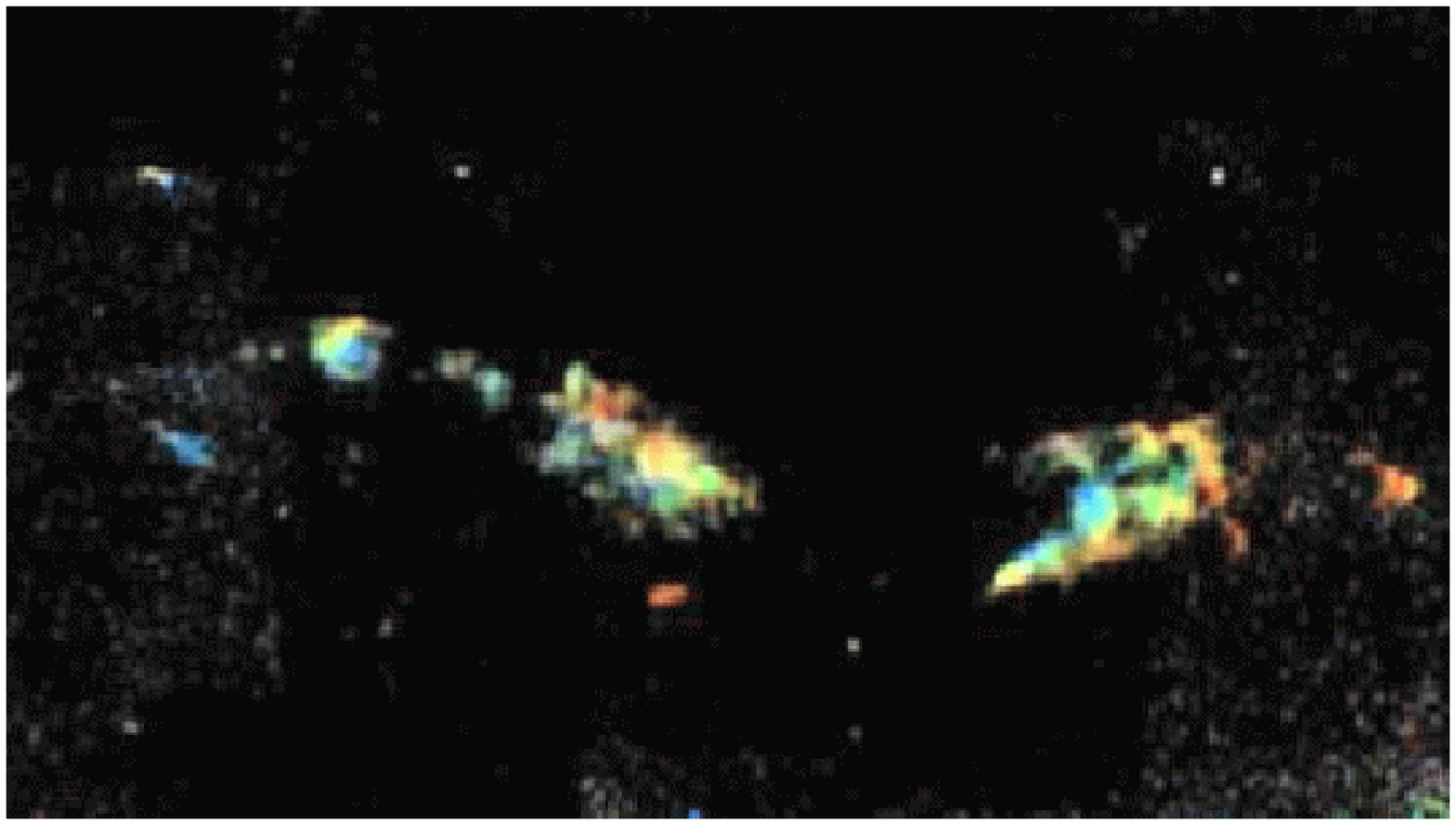}
\end {center}
\caption{Color composite of the doppler shifted H$_2$ line emission
$v$=1-0 $S$(1) in Cepheus A, showing the two main regions Cep~A~(E)
and Cep~A~(W). Colors correspond to velocities in the LSR: Blue
-32.3~km~s$^{-1}$, green -22.5~km~s$^{-1}$, and red
-12.7~km~s$^{-1}$. The region shown is about 5\arcmin$\times$3\arcmin,
North is at the top and East is to the left.
\label{FIG:Doppler_Shift}}
\end{figure*}

  We created a color coded velocity image from the three channel
velocity maps with more copious emission (-32.3, -22.5, and -12.7
km~s$^{-1}$) as blue, green, and red respectively.
Figure~\ref{FIG:Doppler_Shift} presents this color composite map.  It
should be noted that these velocities are somewhat bluer than the
-11.2~km~s$^{-1}$ systemic velocity found from millimeter wavelength
line observations of CO (e.g. \citet{Narayanan96}), as had already
been noted by \citet{Doyon88}.

  The velocity structure in both regions shows a very complex
pattern. However, a slightly systematic change from blue to red
starting at the center of the image can be appreciated on the western
outflow, that is not present in the eastern region. On the other hand,
the Ceph A (E) region shows a large amount of small clumps with
different velocities lying side by side.

\begin{figure*}
\begin{center}
\includegraphics[width=8cm]{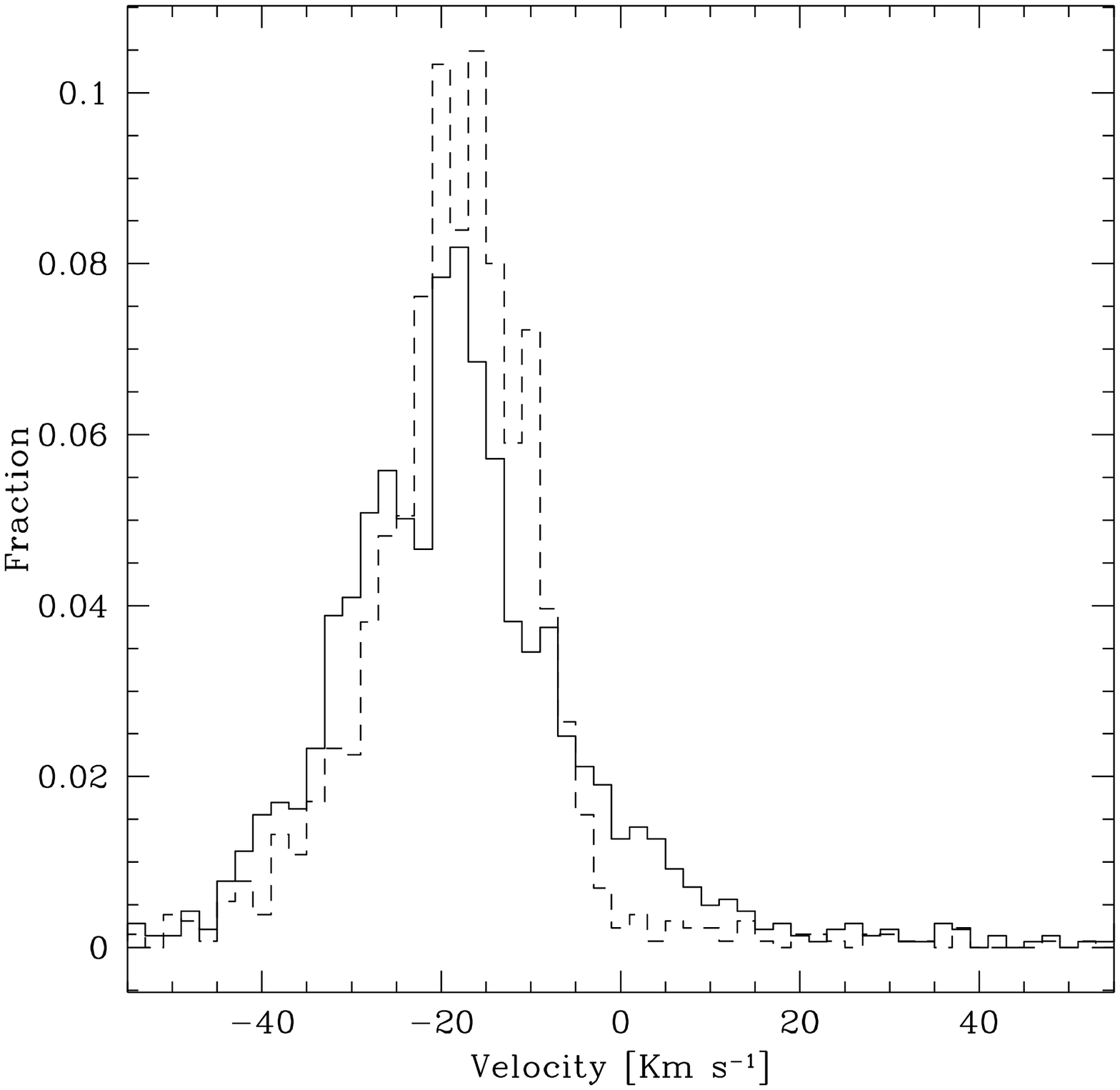}
\end {center}
\caption{Fraction of pixels with a given value of the centroid radial
velocity (LSR) for the Cep~A~(E) ({\it dashed line}) and Cep~A~(W)
({\it solid line}) regions. The size of the bins is 2~km~s$^{-1}$.
\label{FIG:histovel}}
\end{figure*}

  We have calculated the centroid radial velocity for each pixel by
taking only 10 velocity channels around the peak intensity.
Figure~\ref{FIG:histovel} presents histograms of centroid radial
velocity for the two H$_2$ emission regions of the image in
Fig.~\ref{FIG:Doppler_Shift}.  Regardless of the difference in
morphology for the two regions, they have similar fractional
distribution of pixels for a given radial velocity. However, Cep~A~(W)
has a wing that extends into positive radial velocities and a small
peak in the negative velocities.

%figura 2

%figura 3

\begin{figure*}
\begin{center}
\includegraphics[width=\textwidth]{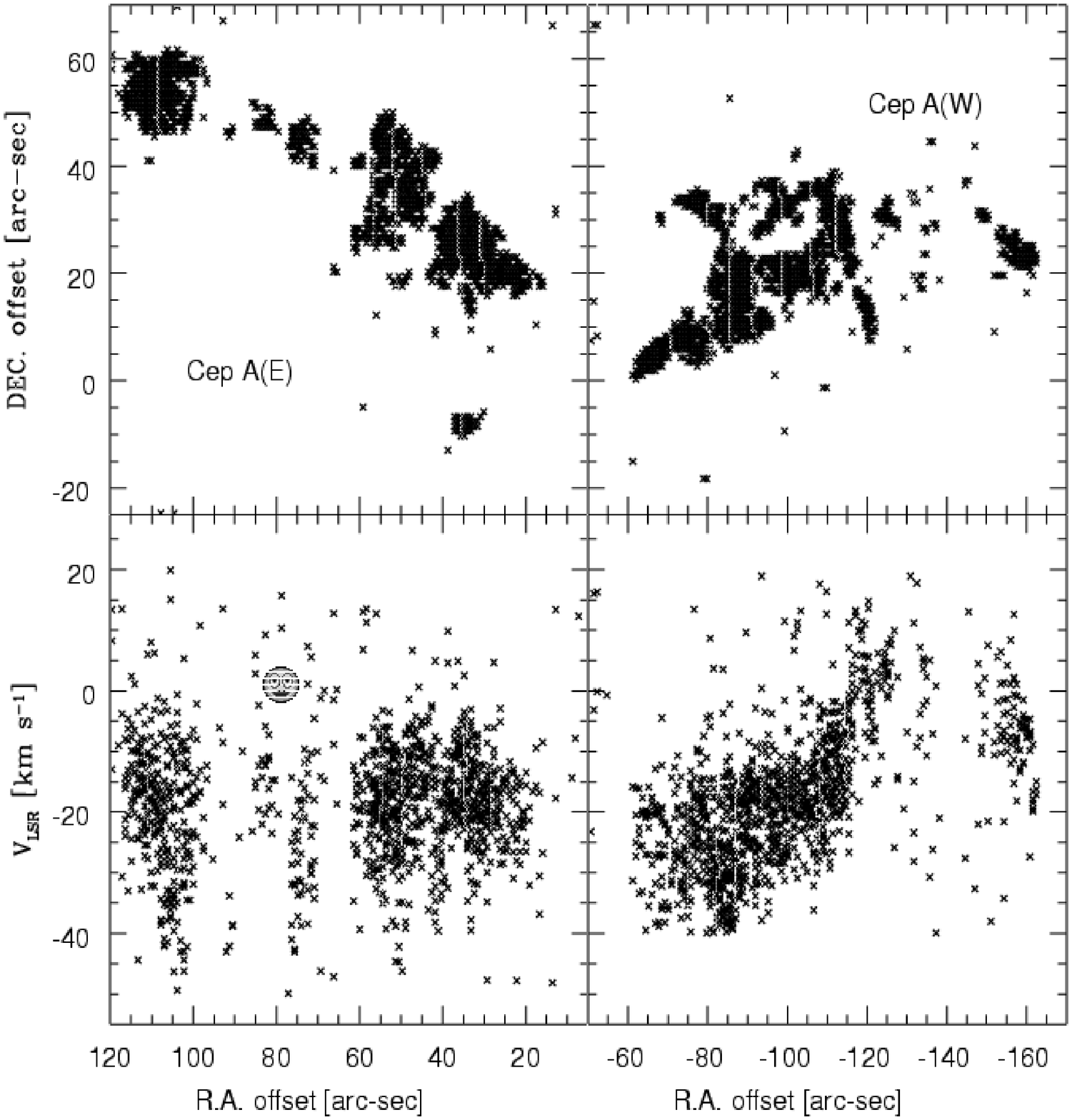}
\end {center}
\caption{Diagrams of detected H$_2$ $v$=1--0 $S$(1) line emission in all
 velocity channels({\it top}) and the velocity-position diagrams
 ({\it bottom}) for Cep~A~(E) ({\it left}) and Cep~A~(W) ({\it
 right}).  Offset values are referred to the position of HW2 (see
 Fig.~\ref{FIG:Channel-map}) considered as position (0,0).
\label{FIG:pos-vel-diag}}
\end{figure*}

  Figure~\ref{FIG:pos-vel-diag} presents the centroid radial velocity
as a function of displacement along the right ascension axis for both
regions Cep~A~(E) and Cep~A~(W). All the channels with detected
emission of H$_2$ from Fig.~\ref{FIG:Doppler_Shift} are shown. The
radial velocity of Cep~A~(E) has an almost constant mean value of
$\sim$-18~km~s$^{-1}$ along the region with a high dispersion around
this value. Meanwhile, radial velocity of Cep~A~(W) increases its
value with offset position to the West. We have fitted a line to each
position-velocity data for each region and found rms residual values
of 8.3 and 10.2 km~s$^{-1}$ for Cep~A(W) and Cep~A(E), respectively.
The smaller velocity dispersion and the large number of small clumps
with different velocities (see Fig~\ref{FIG:Doppler_Shift}) in the East
region indicate a more turbulent outflow in Cep~A(E) compared to
Cep~A(W).

% figura 4

\subsection{A Mach disk in CephA(E)}

\begin{figure*}
\begin{center}
\begin{tabular}{cc}
\includegraphics[width=7cm,height=5.7cm]{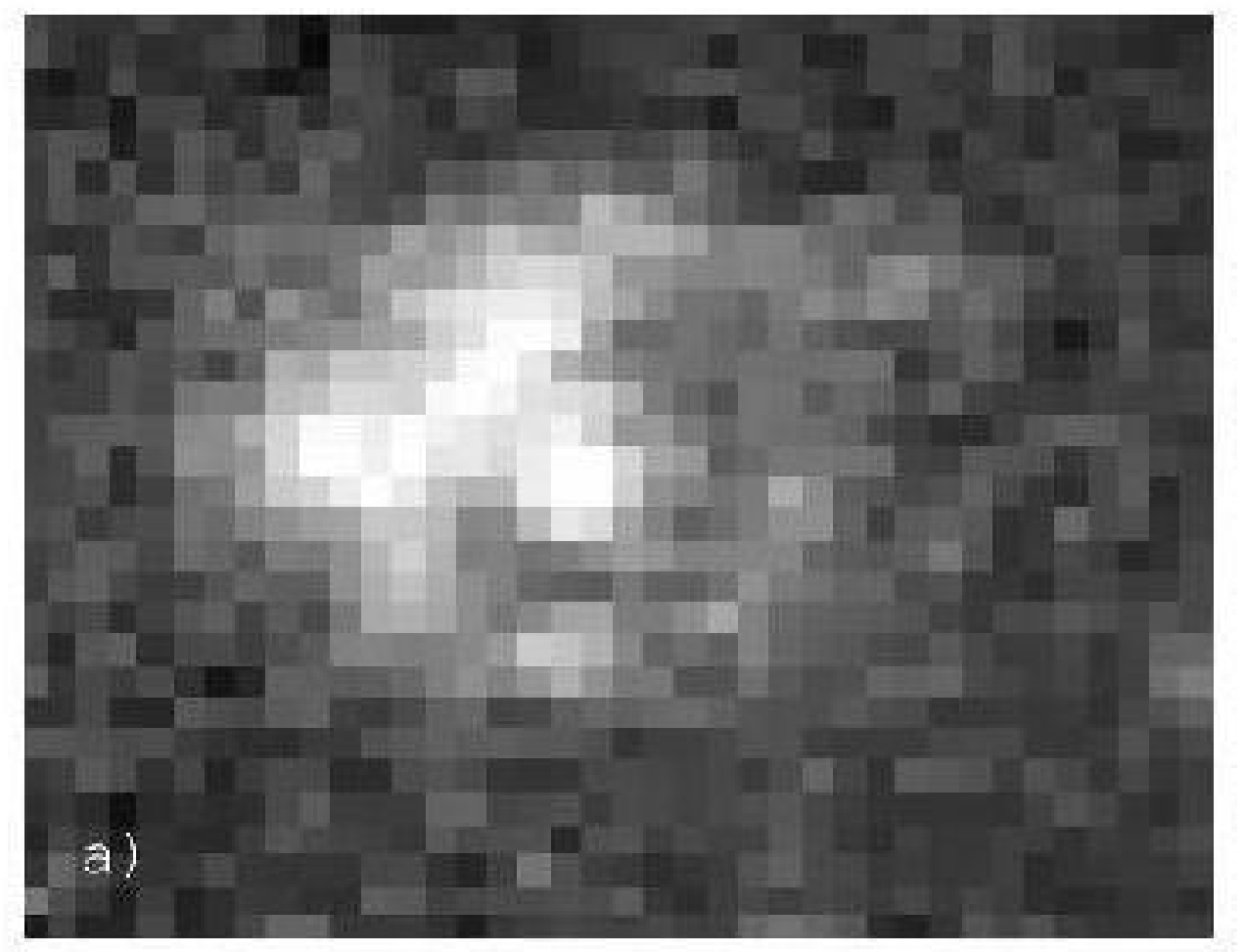} &
\includegraphics[width=7cm,height=5.5cm]{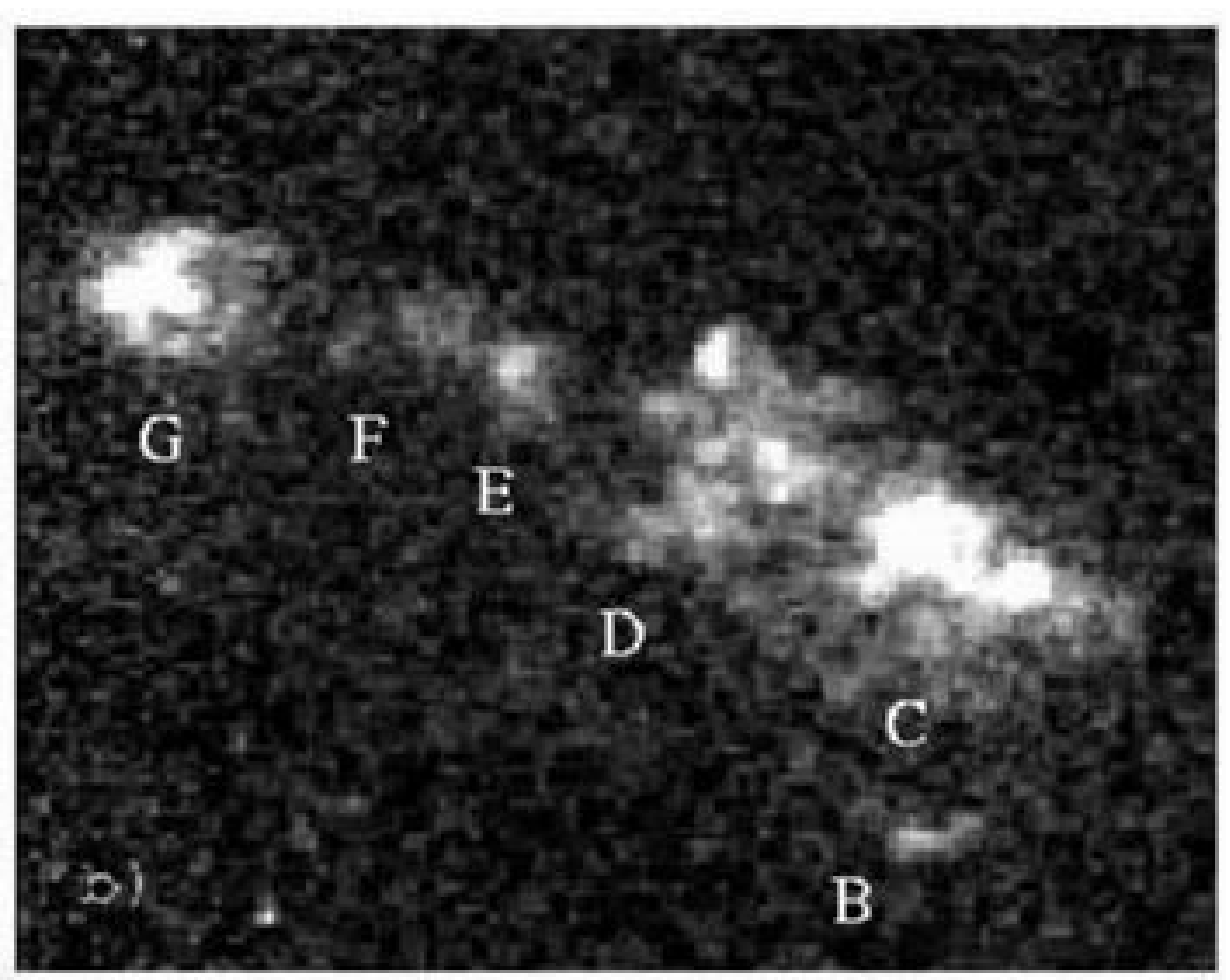} \cr
\includegraphics[width=7cm]{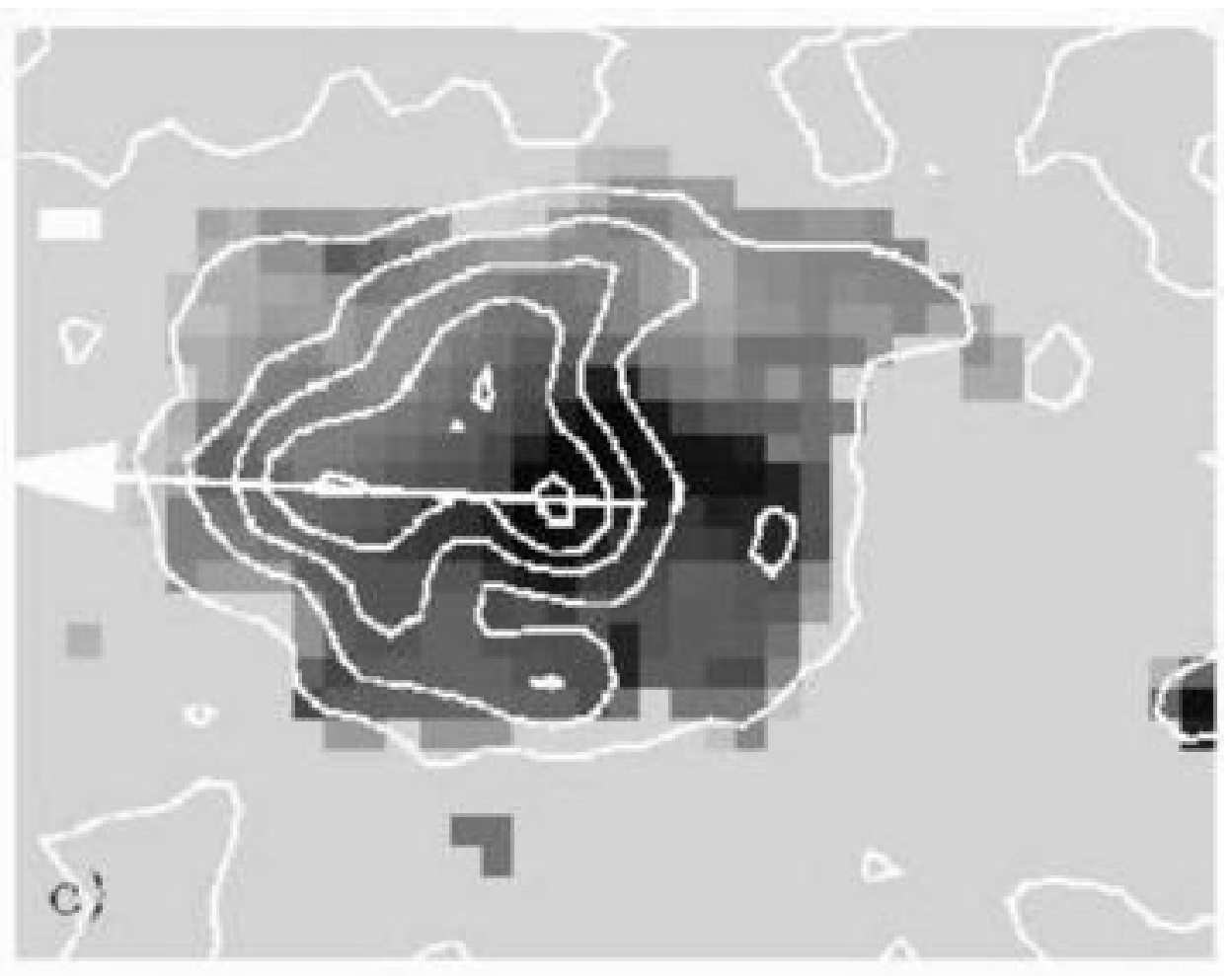} &
\includegraphics[width=7cm,height=5.8cm]{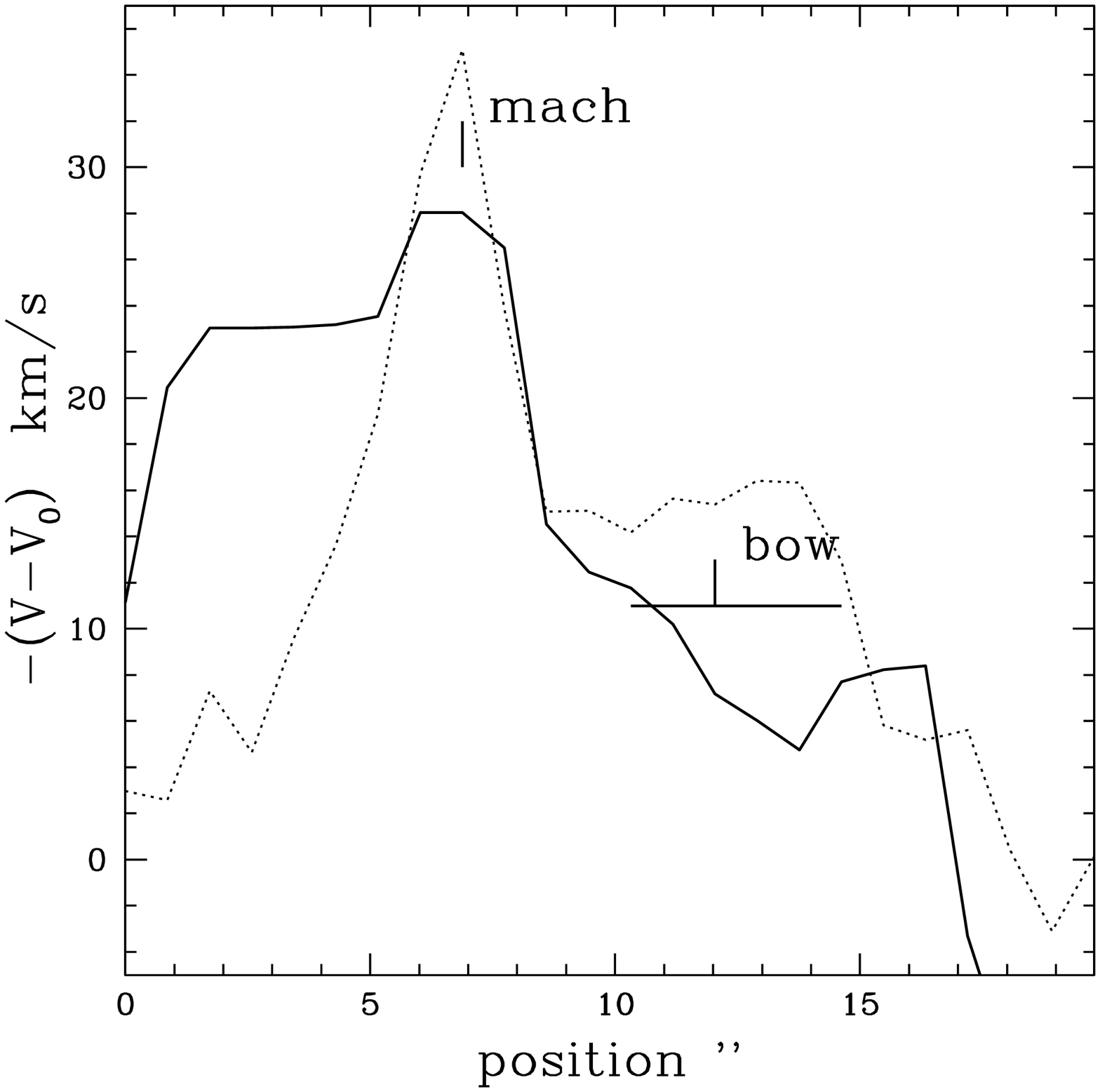}
\end{tabular}
\end {center}
\caption{Mach disk in Cep~A~(E). {\bf (a)} Amplified H$_2$ image of
region G showing the edge of the outflow presented in panel {\bf (b)}.
%indicating the bow shock structure that is amplified in the a) inset.
{\bf (c)} shows the centroid velocity of the bow-shock region G in
gray scale, going from $-40$~km~s$^{-1}$ (black) to 0~km~s$^{-1}$ in
white. Overlaid contours show the intensity map.  {\bf (d)} shows a
position-velocity diagram of a cut shown in {\bf (c)} in the
direction indicated by the arrow. The solid line corresponds to
velocity $-(V-V_0)$, $V_0=-11.3$~km~s$^{-1}$, and the dotted line is 
intensity.
\label{FIG:mach}}
\end{figure*}

  The Cep~A~(E) outflow culminates in an arc shaped structure (labeled
G in Fig.~\ref{FIG:Channel-map}) that resembles a bow shock.  This
region is amplified in Fig.~\ref{FIG:mach}a.  A bright spot can be
seen in the center of the bow. The centroid velocities corresponding
to this region are shown in Fig.~\ref{FIG:mach}c. The highest
blue-shifted velocity of the region (-40~km~s$^{-1}$) corresponds to a
slightly elongated region (in the direction perpendicular to the
outflow) that includes the bright spot and decreases toward the bow,
as can also be seen in a position-velocity diagram in
Fig.~\ref{FIG:mach}d.  This kinematic behavior is expected if the
bright spot corresponds to the Mach disk of the jet, where the jet
material interacts with previously swept material accumulating in
front of the jet and behind the bow shock.

 A few cases are known where a Mach disk is observed in H$_2$.  The
case of \citet{Kumar02} at the N1 outflow in S233IR \citep{Porras00},
where a flattened structure is seen in high definition H$_2$ images,
although it is not supported spectroscopically. In HH 7
\citet{Khanzadyan03} detect a flattened [FeII] structure that
coincides with a blue-shifted knot in H$_2$.  This case is similar to
the present case for Cep~A~(E). Blue-shifted velocity is expected as
the jet material flows in all directions upon interacting at the Mach
disk, in particular toward the present direction of the observer,
nearly perpendicular to the outflow axis. As noted by \citet{Kumar02}
a detection of a Mach disk has rather interesting implications: 1) the
jet would be partially molecular; 2) the velocity of the jet must be
small enough to prevent dissociation of H$_2$ in the Mach disk; and 3)
the jet should be heavy.  In this case we can estimate the velocity of
the molecular jet from the maximum observed velocity (-40~km~s$^{-1}$)
minus the rest velocity of the molecular cloud
($V_0$=-11.3~km~s$^{-1}$) to be around 29~km~s$^{-1}$.  The distance
from the Mach disk to the bow-shock apex is 5\arcsec~ to 8\arcsec~
($\sim$0.016~to 0.025~pc at D=725 pc).

%figura 5

\subsection{Cep~A~(W) velocity gradients}

\begin{figure*}
\begin{center}
\includegraphics[width=\textwidth]{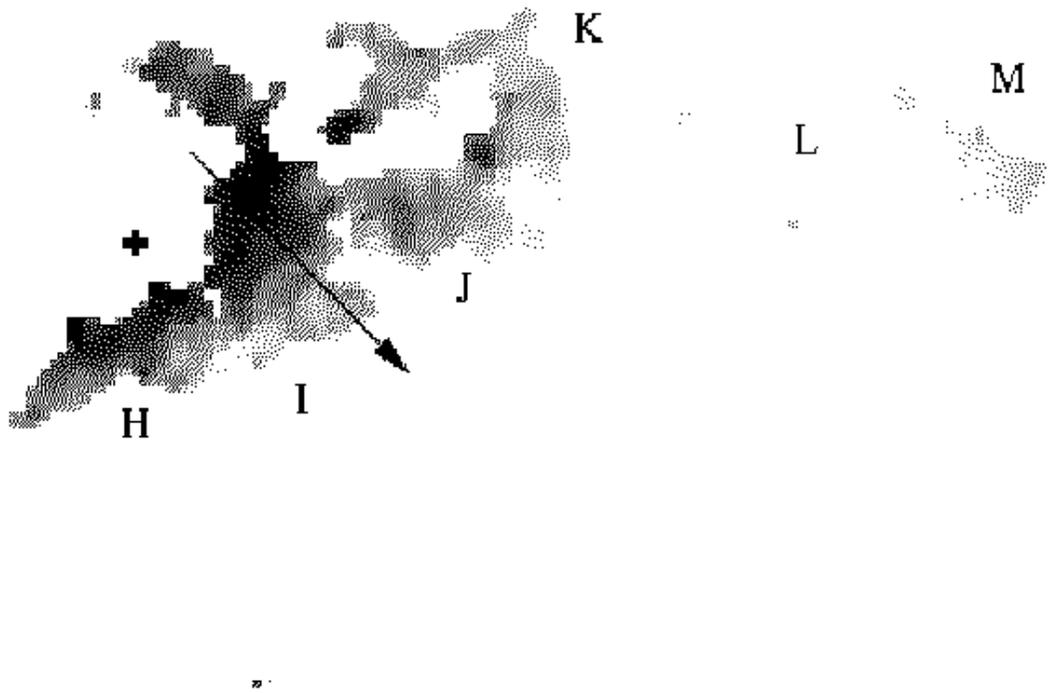}
\end {center}
\caption{Velocity gradients in Cep~A~(W).  The gray levels show
velocity in the range -40~km~s$^{-1}$ (black) to -4~km~s$^{-1}$
(white).
\label{FIG:ladow}}
\end{figure*}

  The western H$_2$ outflow in Cep~A displays a series of wide arcs
reminiscent of thin sections of shells.  The outflow has been
described as a hot bubble \citep{Hartigan00} that drives C-shocks into
the surrounding medium. In some shock fronts along this outflow, they
observed that H$_2$ emission leads the optical [SII]$\lambda$6717
which in turn leads H$\alpha$. This is taken as evidence in favor of a
C-shock that slowly accelerates and heats the ambient medium ahead of
the shock.  We find further evidence of this from the H$_2$
kinematics.  A velocity gradient from higher to lower velocities is
observed in some of the individual arc structures, as is the case for
the one labeled {\bf I} in Fig.~\ref{FIG:Channel-map}.  The spatial
map of the centroid velocities in Cep~A~(W) is shown in
Fig.~\ref{FIG:ladow}.  A smooth velocity gradient, covering from
-36~km~s$^{-1}$ to -8~km~s$^{-1}$, is observed in the SW border of the
arc, in the direction indicated by the arrow.  That is, a velocity
gradient going from blue-shifted to closer to the rest velocity of the
molecular cloud (-11.2~km~s$^{-1}$) in a region of 17\arcsec, or
0.06~pc at D=725~pc.  We regard this as the kinematic evidence of a
C-shock, as the slow acceleration of material ahead of the shock is
quite evident.

%figure 6

\subsection{Flux--velocity Relation}

\begin{figure*}
\begin{center}
\includegraphics[width=7cm]{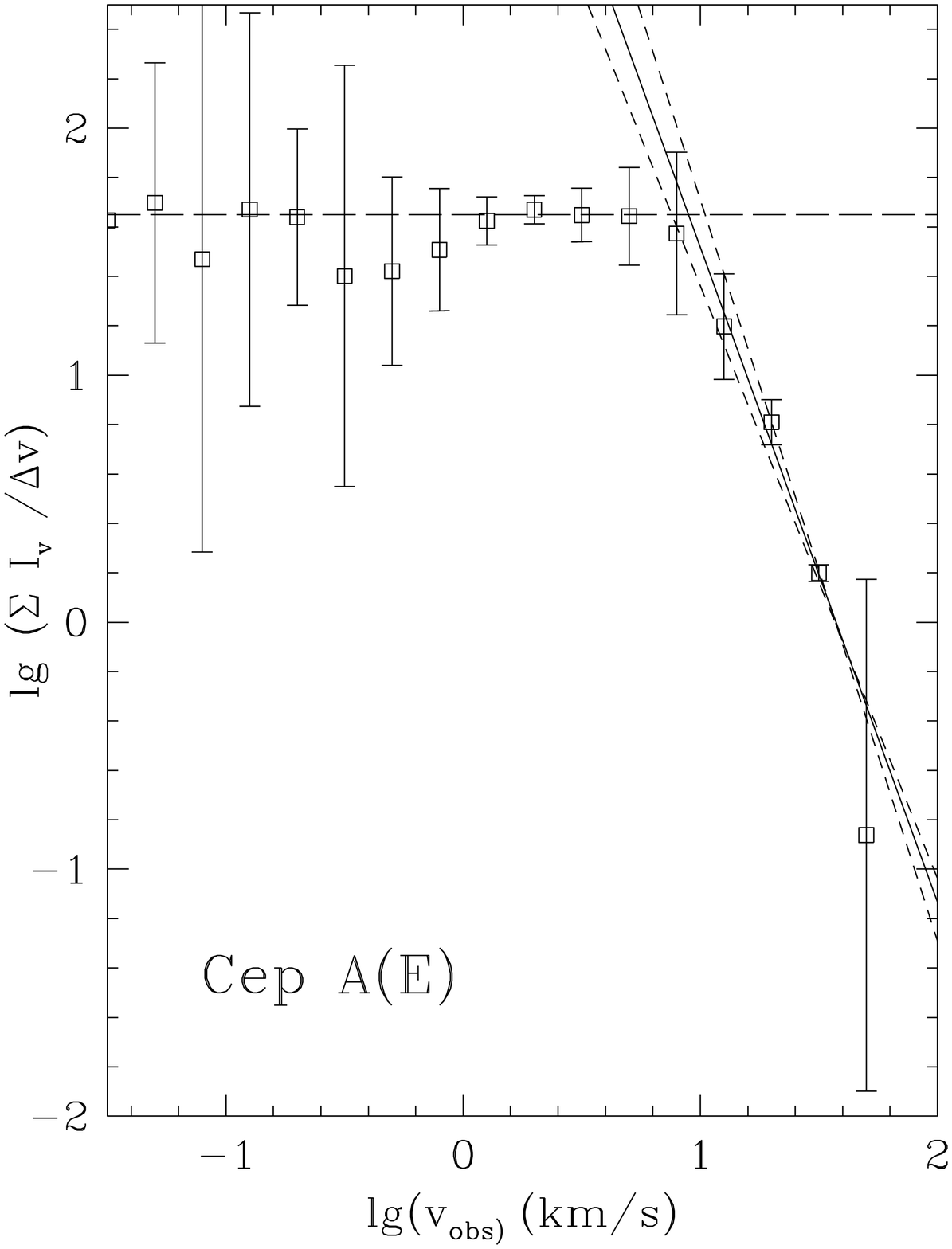}
\end {center}
\caption{Flux-velocity relation of H$_2$ $v$=1--0 $S$(1) line emission
in Cep~A~(E). The long-dashed line shows the flat low velocity
behavior up to a break velocity where the solid line indicates a
power law decrease.  The dashed lines show the range of possible
values of the index (-2.6$\pm$0.3) and of $log(v_{break})$
(0.95$\pm$0.07).
\label{FIG:Flux-vel-diagE}}
\end{figure*}

\begin{figure*}
\begin{center}
\includegraphics[width=7cm]{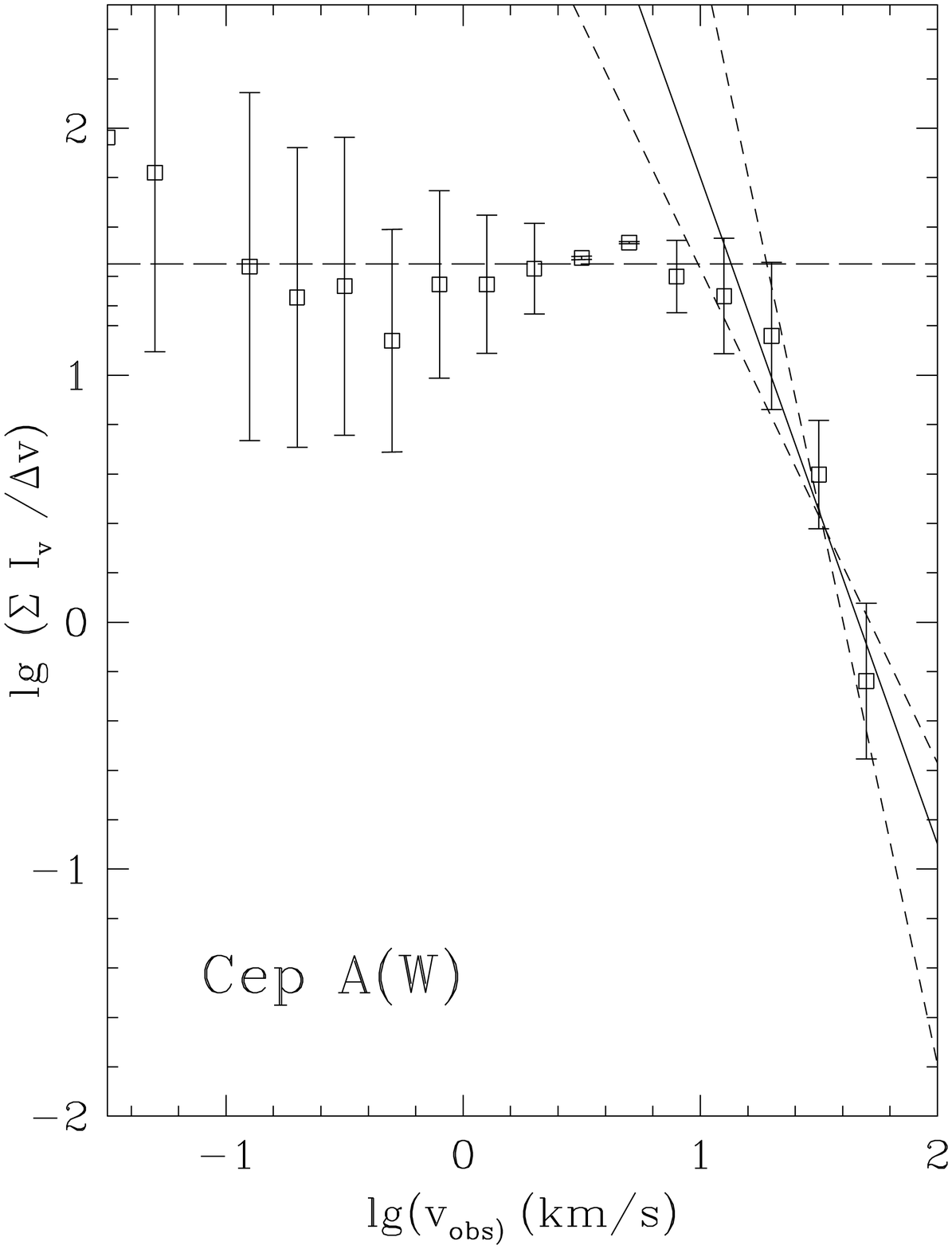}
\end {center}
\caption{Same as in Fig.~\ref{FIG:Flux-vel-diagE}, but for the region
Cep~A~(W). The range of possible values of the power law index is
-2.7$\pm$0.9 and of $log(v_{break})$ is 1.13$\pm$0.13.
\label{FIG:Flux-vel-diagW}}
\end{figure*}

  We have calculated the flux-velocity diagrams separately for the
Cep~A~(E) and Cep~A~(W) outflows, as described in \citet{Salas02}.
For every pixel with signal above the detection threshold, we added
the fluxes of all the pixels with centroid velocities in bins of
observed centroid velocity $\left| v_{obs}\,-\,v_{rest}\right| $, with
respect to the rest velocity of the region (-11.2~km~s$^{-1}$). The
flux-velocity diagrams so obtained are shown in
Fig.~\ref{FIG:Flux-vel-diagE} and \ref{FIG:Flux-vel-diagW}.

  As shown by Salas \& Cruz-Gonz\'alez, this procedure gives similar
flux-velocity relations for a variety of outflows, consisting in a
flat spectrum for low velocities, followed by a power-law decrease
above a certain break-velocity.  The power-law index is very similar
for different outflows, as is the case for Cep~A~(E) and Cep~A~(W)
where it is -2.6$\pm$0.3 and -2.7$\pm$0.9 respectively (solid lines in
figures \ref{FIG:Flux-vel-diagE} and \ref{FIG:Flux-vel-diagW}).  The
break velocity however, is a little different.  The logarithm of
$v_{break}$ (in km~s$^{-1}$) takes values of 0.95$\pm$0.07 and
1.13$\pm$0.13 respectively, a difference of around 2$\sigma$ which
suggests that $v_{break}$ may be larger for Cep~A~(W).  As was
discussed in \citet{Salas02}, outflows of different lengths ($l$) show
break-velocities varying as $v_{break}\propto l^{0.4}$, a result that
is taken to imply an evolutionary effect, similar to the case of CO
outflows \citep{Yu99}.  However, in the case of Cep~A~(E) and
Cep~A~(W) the outflow length is very similar, as might be the outflow
age.  Salas \& Cruz-Gonz\'alez also argue that other causes for a
difference in break velocities could be the amount of turbulence in
the outflow, which might be the case for Cep A as is mentioned at the
end of \S 3.1.  We will next explore this possibility through the use
of a wavelet analysis.

%figure 7

%figure 8

\section{Wavelet Analysis of the H$_2$ emission from Cep~A}
\label{sec:Wavelet}
\subsection{Description}

  Both H$_2$ emission regions in Cep~A show a rather complex
velocity-position structure (see Fig.~\ref{FIG:Doppler_Shift}).  We
have carried out a wavelet analysis in an attempt to understand the
relation between sizes of clumps, velocity, and velocity dispersion as
a function of position along the outflows.

  The wavelet transform analysis has been used before by
\citet{Riera03} to analyze H$\alpha$ observations using a
Fabry-P\'erot of the complex outflow in the HH~100 jet.  Although,
their procedure is equivalent to the one followed by \citet{Gill96} to
study turbulence in molecular clouds, we have followed the analysis by
\citet{Riera03} because it is more relevant to the study of outflows.

  The two-dimensional wavelet transform analysis is used to obtain
the positions of all H$_2$ emission clumps as well as their
characteristic sizes.  The sizes of these clumps are determined from
total H$_2$ flux images created by adding the continuum subtracted
intensities of all the channel maps.  Once obtained the positions and
characteristic sizes of the H$_2$ structures of Cep~A, the spatial
averages (over the characteristic sizes of the structures) of the line
center velocity and of the line widths are calculated to study the
turbulence in the two regions of Cep~A.

  Our case is similar to \citet{Riera03} in the sense that Cep~A
regions have an axial symmetry and that the kinematics properties seem
to depend on the offset along this axis. Then, it is reasonable to
expect that the characteristic size may be different along and across
the regions. To exploit this symmetry, we rotated the H$_2$ image so
the long part of both regions, the main axis, is parallel to the $x$
coordinate defined along the east-west direction.  On this rotated
images, we carried out an anisotropic wavelet decomposition using two
bases of different sizes, one along and one across the main axis.

  We adopted a basis of ``Mexican hat'' wavelets of the form
\begin{equation}
g(r;a_x,a_y) = C (2-r^2)e^{-r^2/2}   \; ,
\end{equation}
where $r=[(x/a_x)^2+(y/a_y)^2]^{1/2}$; $a_x$ and $a_y$ are the scale
lengths of the wavelets along the $x$- and $y$-axes, respectively, and
$C = (a^2_x + a^2_y)^{-1/2}$. This is a very common used wavelet, but
we also choose to use it because simplifies the detection of intensity
peaks,  approaches better to the shape of the intensity peaks, and
behaves well under FFT calculations.

  To compute the wavelet transform we have to calculate the convolutions
\begin{equation}
T_{a_x,a_y}(x,y) = \int \int I(x',y')g(r';a_x,a_y)dx'dy'   \; ,
\end{equation}
for each pair ($a_x,a_y$), where
$r'=\{[(x'-x)/a_x]^2+[(y'-y)/a_y]^2\}^{1/2}$, $I(x,y)$ is the
intensity at pixel position $(x,y)$, and $g(r';a_x,a_y)$ is give by
eq. (1). These convolutions are calculated by using a Fast Fourier
Transform (FFT) algorithm \citep{Press92}.

  The wavelet transformed images $T_{a_x,a_y}(x,y)$ correspond to
smoothed versions of the intensity of the H$_2$ image. We use this
images to find the sizes of the structures in the H$_2$ regions of
Ceph~A. First, on the transformed image with $a_x=a_y=1$, we fixed
the position of $x$ and find all the values of $y$ where
$T_{a_x,a_y}(x,y)$ has a local maximum. Several maxima may be found
for each position $x$ that correspond to different structures observed
across the regions. The maxima found with $a_x=a_y=1$ will correspond
also to the local maxima of $I(x,y)$.

 For each pair ($x$,$y$) where $I(x,y)$ has a maximum we determine
$(a_x,a_y)$, in the $a_x$ and $a_y$ space, where the wavelet transform
has a local maximum.  $a_x$ and $a_y$ will be the characteristic size
of the clump with a maximum intensity at ($x$,$y$). The ($a_x$,$a_y$)
space is search in such way that we first identify the size of the
smaller clumps and then the bigger ones. This progressive selection
allow us avoid to choose clumps that overlap with its
neighbors. Naturally, the biggest clumps will have an structure
similar to the whole region.

\subsection{Size of the H$_2$ Clumps}
\label{subsec:Size}
\begin{figure*}
\begin{center}
\includegraphics[width=\textwidth]{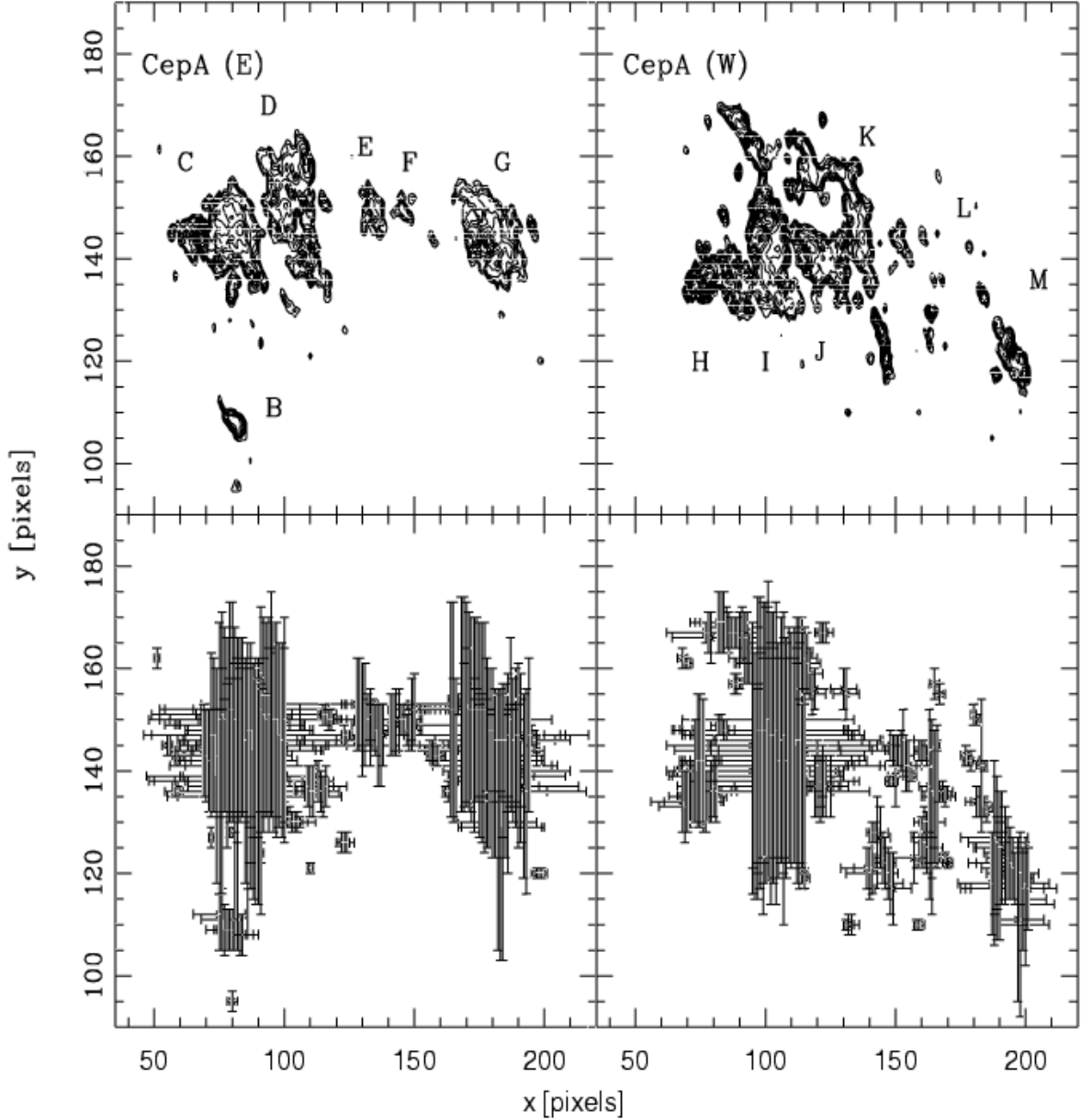}
\end {center}
\caption{ Position and characteristic sizes of the structures ({\it
 bottom}) and contour plots of the H$_2$ image ({\it top}) obtained
 from the velocity channels for Cep~A~(E) ({\it left}) and Cep~A~(W)
 ({\it right}). Contour levels shown from 0.0 to 1.5~counts/sec in
 0.1~counts/sec increments for both plots. The plus sign shows the
 positions of the maximal $y_k$ of the wavelet transform obtained for
 the different values $x$ along the region. The characteristic sizes
 $a_{x,k}$ and $a_{y,k}$ of these maximal are shown as error bars
 centered on the positions of the maximal. We have separately rotated
 the map of each region by 19$^\circ$ so they are more or less
 parallel to the $x$-axis.  The distances $y$ (across the regions)
 and $x$ (along the regions), in pixels units, are with respect to the
 rotated image and have an arbitrary origin. In the Cep~A~(E) case the
 $x$ axis has been inverted (West is to the right) for analysis
 purpose (see text).
\label{FIG:Size}}
\end{figure*}

  The results obtained with the process described above are shown in
Fig.~\ref{FIG:Size}.  This figure shows the two images of molecular
hydrogen emission of Cep~A, which have been rotated by 19$\degr$ so
the longways dimension of the regions are more or less parallel to the
$x$-axis. In the case of Cep~A~(E), the $x$ axis has been also
inverted (West is to the right) so the $x$ values in both panels in
Fig.~\ref{FIG:Size}, although arbitrary in origin, are an estimate of
the offset from the central region between the two regions. It has to
be clarified that the $x$ coordinates are values from independent
images so there is really no correlation between them.  However, the
span in the vertical and horizontal axis are kept the same in both
panels of Fig.~\ref{FIG:Size} to ease the size comparison for each
region. Six large structures (B to G) were identified in Cep~A~(E) and
six (H to M) for Cep~A~(W) (see Fig.~\ref{FIG:Channel-map}). The
spatial limits of these regions are shown in Table~\ref{TAB1}.

\begin{table*}
\caption{ H$_2$ Emission Regions in Cep~A \label{TAB1} } 
\begin{tabular}{ccccccc}
\tableline\tableline
Region  & $x_{min}$ & $ x_{max} $ & $y_{min}$ & $y_{max}$ & $m$  &$\alpha$\\
\tableline
  B     &   72      &    88       &  102      &  115      & 0.612 & -0.073:\\
  C     &   53      &    91       &  129      &  159      & 0.774 & 0.140 \\
  D     &   92      &   119       &  133      &  170      & 0.702 & 0.169 \\
  E     &   127     &   140       &  142      &  158      & 0.739 & 0.155: \\
  F     &   140     &   153       &  144      &  155      & 0.394 & 0.644 \\
  G     &   163     &   199       &  132      &  160      & 0.768 & 0.249 \\
  H     &   67      &   92        &  129      &  144      & --    & 0.397: \\
  I     &   92      &   110       &  129      &  155      & 0.871 & -0.089:\\
  J     &   113     &   129       &  132      &  147      & 0.654 & 0.563 \\
  K     &   130     &   142       &  115      &  162      & --    & 0.379 \\
  L     &   142     &   187       &  115      &  150      & 0.931 & 0.439 \\
  M     &   188     &   206       &  113      &  131      & 0.937 & 0.376 \\ 
\tableline
\end{tabular}

\tablenotetext{}{Col.(1).-- Region B--G Cep~A(E); Region H--M Cep~A(W) (see
Fig.~\ref{FIG:Size}). }

\tablenotetext{}{Col.(2)--(5).-- $x_{min}$, $x_{max}$, $y_{min}$, and
$y_{max}$ are the  boundary coordinates (in pixels) of the region (see
Fig.~\ref{FIG:Size}). }

\tablenotetext{}{Col.(6).-- $m$ is the slope of the linear fit $a_y=ma_x+b$,
where $a_x$ and $a_y$ are the sizes of the identified H$_2$ structures on
the region (see \S~\ref{subsec:Size})}

\tablenotetext{}{Col.(7).-- $\alpha$ is the slope of the linear fit $
\langle \Delta v^2 \rangle ^{1/2} = \alpha (a_x^2
+a_y^2)^{1/2}+\beta$, where $ \langle \Delta v^2 \rangle ^{1/2}$ is
the rms velocity dispersion on the structure of size $a_x$ and $a_y$ (see
\S~\ref{subsec:Deviations})}

\end{table*}

  The H$_2$ intensity maps have then been convolved with a set of
wavelets $g(r;a_x,a_y)$ with $1 \leq a_x \leq 30$ pixels and $1 \leq
a_y \leq 30$~pixels for Cep~A~(E) and $1 \leq a_x \leq 35$ pixels and
$1 \leq a_y \leq 35$~pixels for Cep~A~(W), with a resolution of
0.853~arc-sec per pixel in the $x$-direction and 0.848~arc-sec per
pixel in the $y$-direction. The upper limiting values for $a_x$ and
$a_y$ were selected to allow very few overlapping in the size of the
regions in adjacent peaks.

  The values of the peaks $y_k$ of the wavelet transform obtained for
each position $x$ along the region are shown in the bottom panel of
Fig.~\ref{FIG:Size}. Also shown as error bars are the values of
$a_{x,k}$ and $a_{y,k}$ corresponding to each peak, which give an
estimate of the characteristic sizes along and across the observed
structures.

%figure 9

\begin{figure*}
\begin{center}
\includegraphics[width=\textwidth]{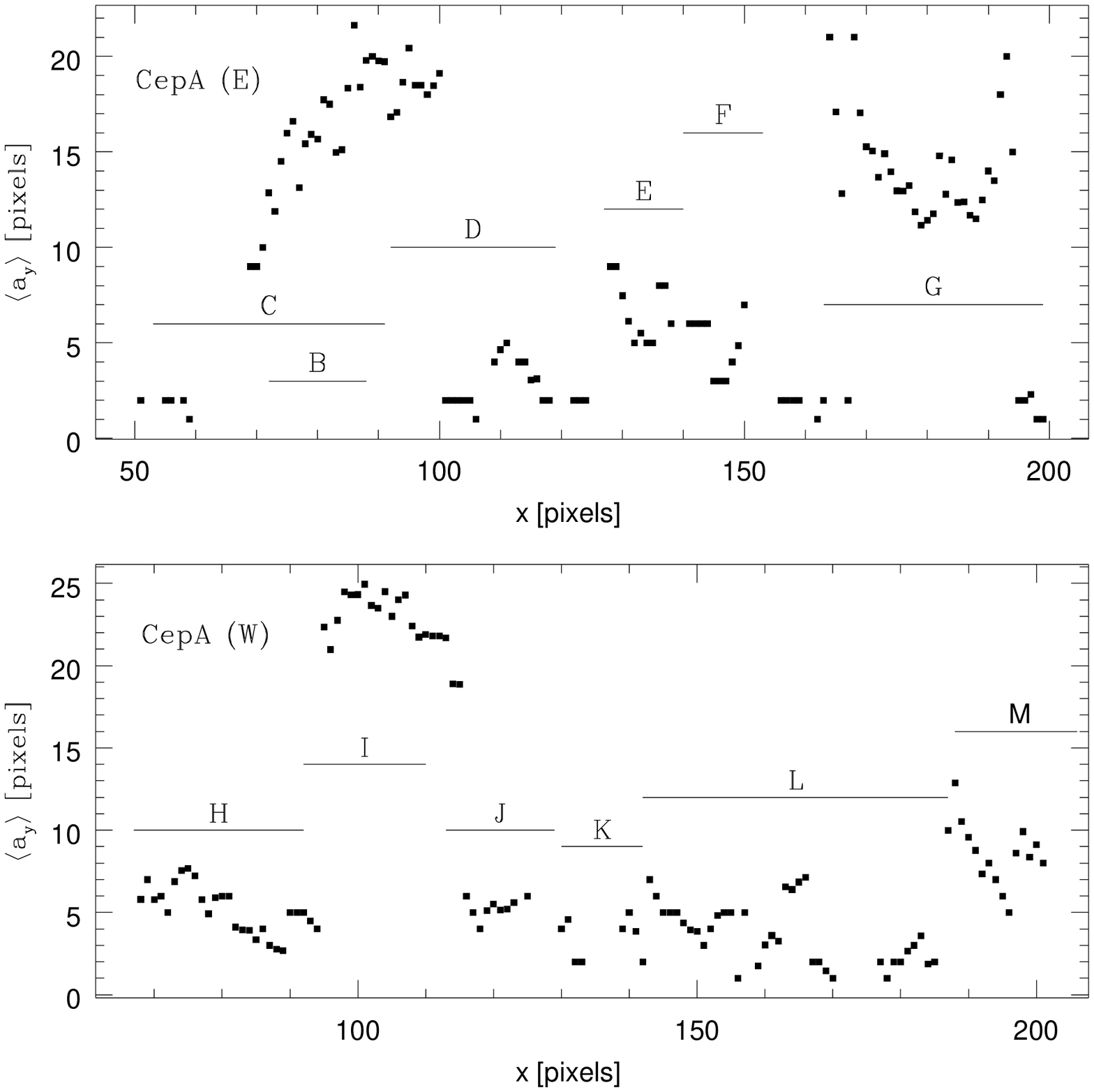}
\end {center}
\caption{Values of $\langle a_y \rangle $ plotted as a function of
position $x$ along the Cep~A~(E), {\it top} panel, and Cep~A~(W) at
the {\it bottom} one.
\label{FIG:broad}}
\end{figure*}

  To quantify the observed broadening of the region, for each position
$x$ we compute the weighted mean of the $y$-spatial scale
perpendicular to the longest axis of the regions:
\begin{equation}
\langle a_y \rangle = \frac{\sum _k a_{y,k}
         T_{a_{x,k},a_{y,k}}(x,y_k)} {\sum _k
         T_{a_{x,k},a_{y,k}(x,y_k)}} \; .
\end{equation}
Figure~\ref{FIG:broad}~ shows $ \langle a_y \rangle$ as a function of
position $x$ along the two Cepheus~A regions.  In both outflows we notice
the presence of small (2 pixels) and large (25 pixels) structures
located with no discernible order along the axial position.

%figura 10

\begin{figure*}
\begin{center}
%\plottwo{Eaxay.eps}{Waxay.eps}
\begin{center}
\begin{tabular}{cc}
\includegraphics[bb=230 0 470 700,width=7cm,height=19cm]{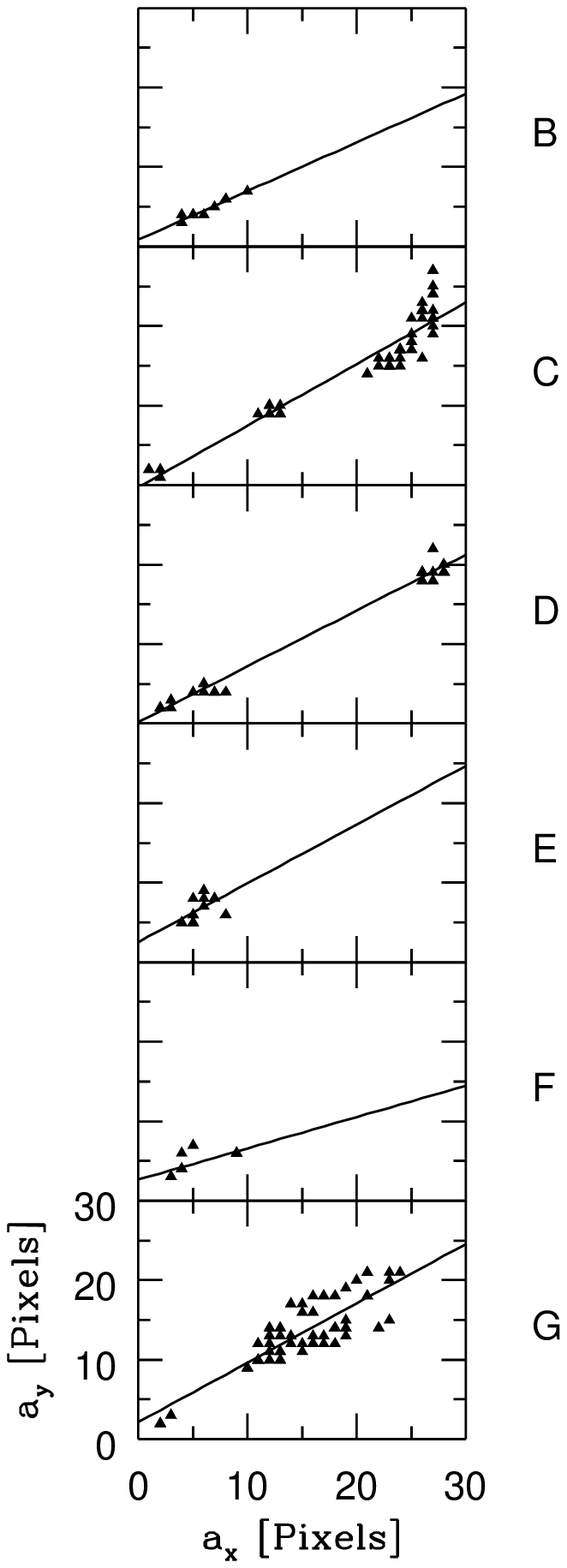} & 
\includegraphics[bb=230 0 470 700,width=7cm,height=19cm]{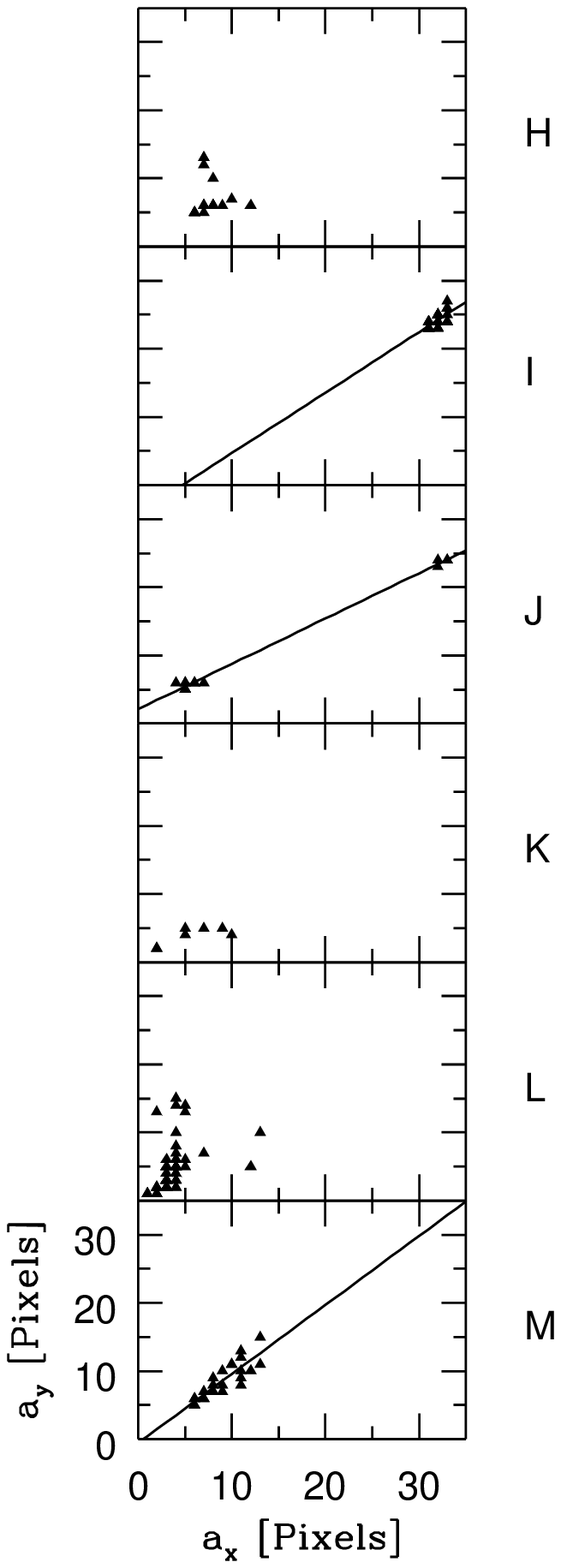}
\end{tabular}
\end{center}
\caption{{\bf Left:} Values of $a_{y,k}$ plotted as a function
$a_{x,k}$ for the different regions along Cep~A~(E).  The linear fits
to the points are drawn as solid lines. The slope $m$ of the linear
fit is listed in Table~\ref{TAB1}. {\bf Right:} Same but for the
region Cep~A~(W).}
\label{FIG:Eaxay}
\end{center}
\end{figure*}

  The values of $a_{y,k}$ as a function of $a_{x,k}$ are shown in
Fig.~\ref{FIG:Eaxay} Cep~A~(E) and Cep~A~(W). These graphs represent a
measure of the symmetry of the structures.  Isotropic structures are
expected to attain $a_{y,k} = a_{y,k}$ and should thus be located on a
line of unitary slope. The values of the slope $\alpha$ for each
region are also given in Table~\ref{TAB1}. We note that all the values
are in general less than one, indicating structures that are longer
than wider.  The western outflow, however, shows relatively wider
structures than Cep~A~(E), a property which has also been
qualitatively observed. It may be possible that the individual
structures follow the pattern of each region as a whole, with the
eastern region being longer than wider while the western region is
the opposite. However, the nature of these anisotropies it is not
clear.

%figura 11

\subsection{Spatial distributions of the radial velocities
and the line widths} 

  In this section we describe the spatial dependence of the kinetic
properties of the two main H$_2$ emission regions of Cep~A.

 From the cube of position-velocity data, we calculate two moments of
the line profiles for each pixel:
\begin{equation}
V_c = \frac{\int v I_v dv}{\int I_v dv}   \; ,
\end{equation}
and
\begin{equation}
W^2 = \frac{\int (v-V_c)^2 I_v dv}{\int I_v dv}   \; .
\end{equation}
In these equations, $v$ is the radial velocity, and $I_v$ the
intensity at a fixed position $(x,y)$ of successive channel maps.  The
integrals are carried out over all the velocity channel maps. Here
$V_c$ is the barycenter of the line profile (i.e. the ``line center''
radial velocity), and $W$ is a second-order moment that reflects the
width of the line profile.

  With these values of $V_c$ and $W$ computed for all positions
$(x,y)$ on the plane of the sky, we calculate the following spatial
averages \citep{Riera03}:
\begin{equation}
\langle V_c \rangle = \frac{\int _{S_{a_x,a_y}} V_c(x',y') I(x',y') dx'dy'}
{\int _{S_{a_x,a_y}} I(x',y') dx'dy'}  \; ,
\end{equation}

\begin{equation}
\langle W^2 \rangle = \frac{\int _{S_{a_x,a_y}} W^2(x',y') I(x',y') dx'dy'}
{\int _{S_{a_x,a_y}} I(x',y') dx'dy'}  \; ,
\end{equation}

\begin{equation}
\langle \Delta v^2 \rangle = \frac{\int _{S_{a_x,a_y}} [V_c(x',y')- \langle V_c \rangle]^2 I(x',y') dx'dy'}
{\int _{S_{a_x,a_y}} I(x',y') dx'dy'}  \; ,
\end{equation}
where $I(x',y')$ is the H$_2$ flux obtained from co-adding all the
channel maps.  These integrals are carried out over areas $S_{a_x,
a_y}$ which are ellipses with central positions ($x,y$) and major and
minor axes, $a_x$ and $a_y$, corresponding to all the values that have
been identified as peaks of the wavelet transform.

  The value of $\langle W^2 \rangle(x,y)$ corresponds to the line
width spatially averaged over the ellipse $S_{a_x a_y}$ with a weight
$I(x',y')$. The same weight spatial average is calculated for the line
center velocities $ \langle V_c \rangle$ within the ellipse, as well
as the standard deviation $ \langle \Delta v^2 \rangle^{1/2}$ of these
velocities.

\begin{figure*}
\begin{center}
\includegraphics[width=\textwidth]{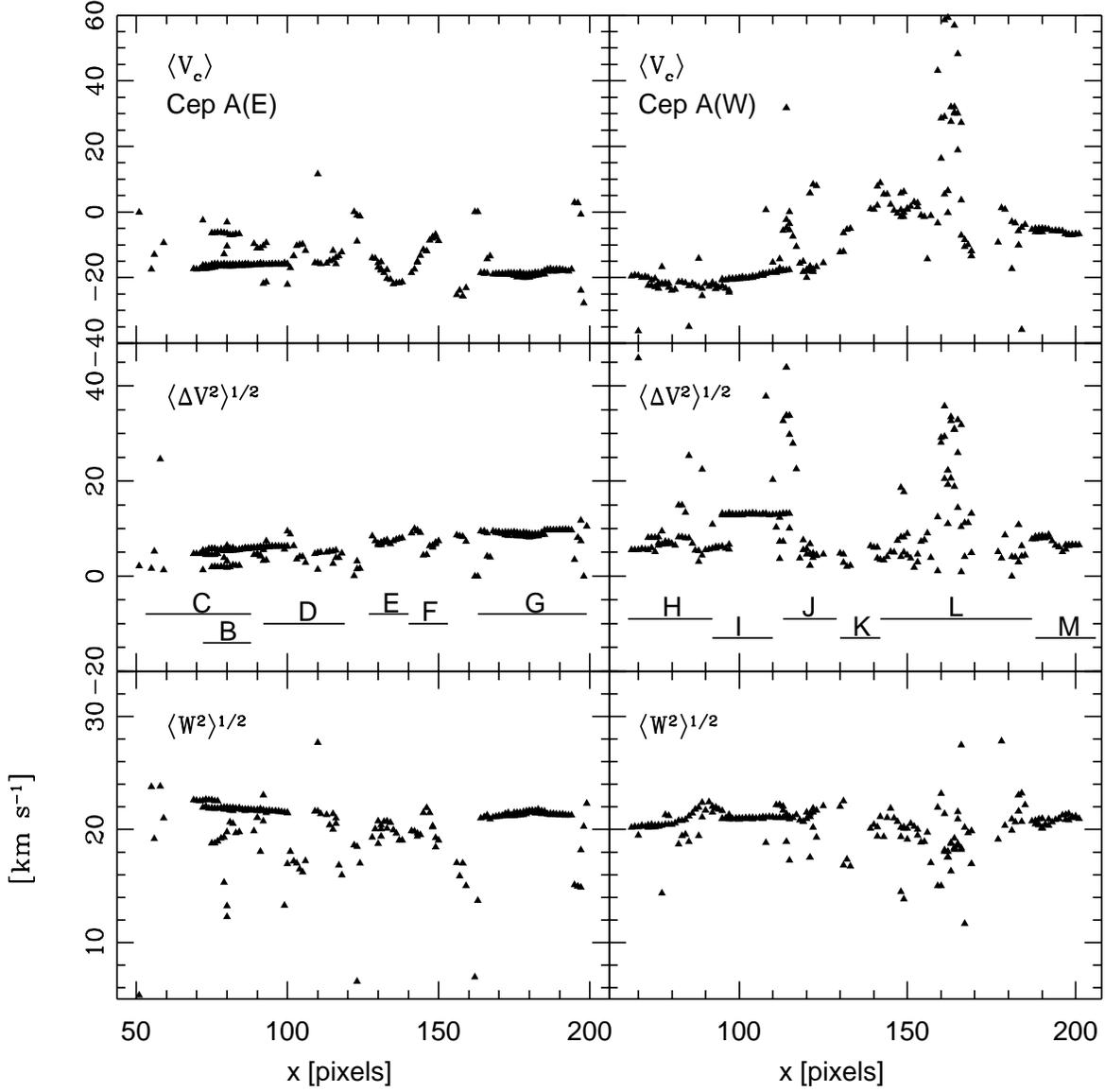}
\end {center}
\caption{Average line centers ({\it top panels}), standard deviations
({\it middle)}, and line widths ({\it bottom}) as a function of
position $x$ along the regions Cep~A~(E) ({\it left column}) and
Cep~A~(W) ({\it right column}). All points with different positions
$y$ across the region and with different $a_x$ and $a_y$ are plotted.
\label{FIG:kine}}
\end{figure*}

  Figure~\ref{FIG:kine} shows the line centers, widths, and standard
deviations as a function of position $x$ along the region (all of the
points at different positions $y$ across the region and with different
$a_x$ and $a_y$ are plotted). The line center or centroid velocity
displayed in the top panels, shows different behaviors for Cep~A~(E)
and Cep~A~(W).  As had been previously noted, Cep~A~(E) has a constant
velocity along the outflow ($\sim$-19~km~s$^{-1}$) while Cep~A~(W)
shows a velocity gradient from about -21 to -2 ~km~s$^{-1}$.  A large
dispersion in the line centers for Cep~A~(W) around positions 115 and
162 is probably due to insufficient S/N, since a similar increase in
the dispersion velocity (medium panel) is concurrent, but lacks a
corresponding increment of the line width (bottom panel).  Other
features present in these figures seem real.  Most remarkably, the
velocity dispersion (medium panel) for Cep~A~(E) increases
monotonically with distance from the source, going from around 4 to
10~km~s$^{-1}$ in 150 pixels or 0.5 pc. No such increase is observed
in the western outflow, where we obtain a constant 3$\pm$2~km~s$^{-1}$
velocity dispersion.  The line widths (bottom panel) seem dominated by
the width of the instrumental profile of 22~km~s$^{-1}$.

   Hence, we conclude from this analysis that the eastern and western
outflows seem intrinsically different.  While the eastern outflow
shows a constant line center velocity and an increasing velocity
dispersion, the western outflow behaves otherwise, showing a velocity
gradient and a constant velocity dispersion with a lower value.  This
had been noticed in our qualitative analysis of the observations. Now,
if we take the velocity dispersion within each cell as a measure of
turbulence, which seems reasonable, then the region Cep~A~(E) is more
turbulent than Cep~A~(W), and it is noticeable that the turbulence
increases with distance from the ``central'' source compared to a
constant behavior in the western source.

% figura 12

\subsection{Deviations of the line center velocity and the size of the region}
\label{subsec:Deviations}

\begin{figure*}
\begin{center}
\includegraphics[width=\textwidth]{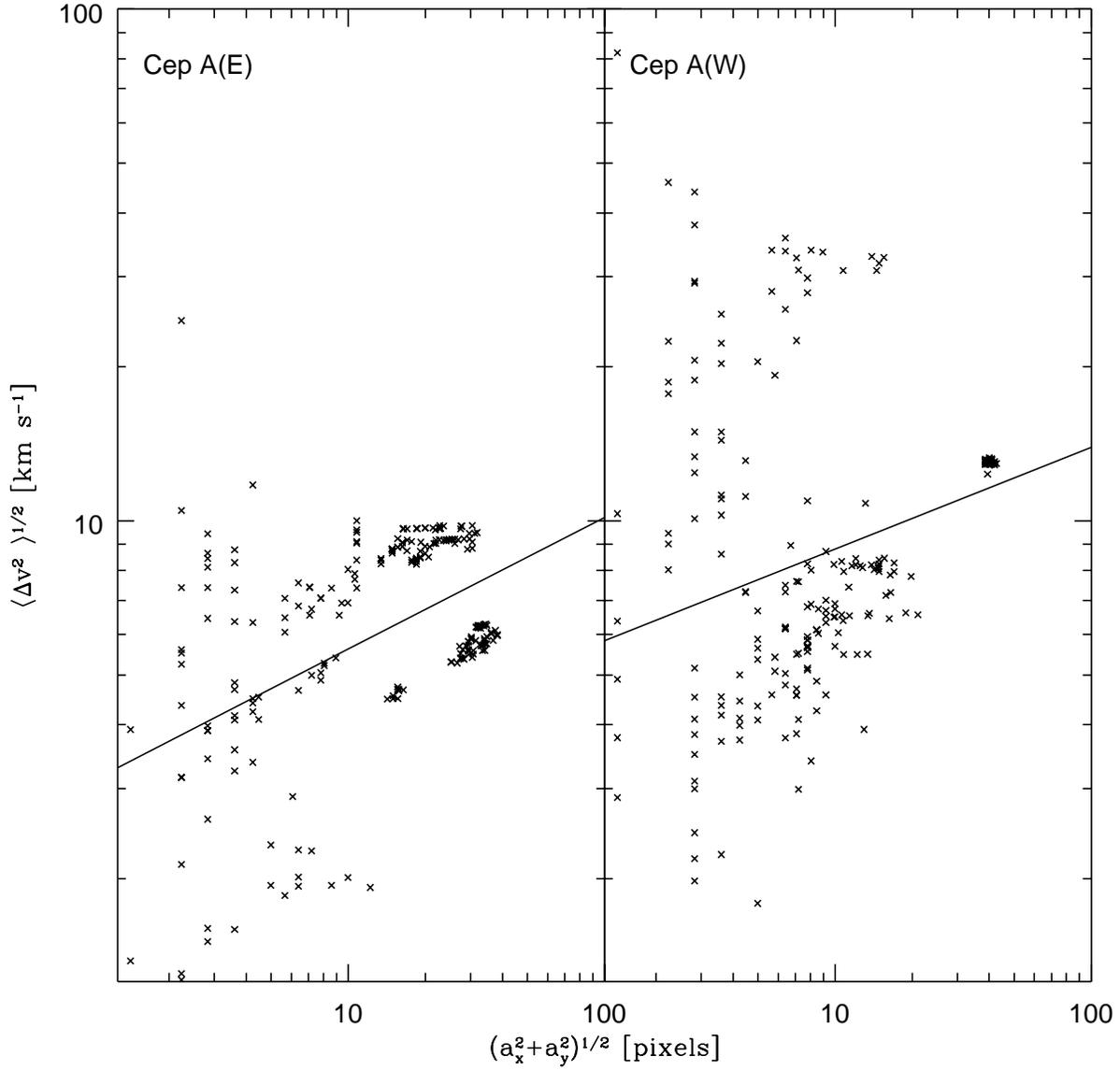}
\end {center}
\caption{ Relation of the deviations of the line center velocity, $
\langle \Delta v^2 \rangle ^{1/2}$ (or $\sigma$), averaged over
regions chosen from the wavelet spectrum, and the size $a$ of the
region, $(a_x^2 +a_y^2)^{1/2}$, for the complete Cep~A~(E) ({\it
left}) and Cep~A~(W)({\it right}) regions. The solid line is the best
fit to the points with a slope $\alpha$ of 0.2574 for Cep~A~(E) and
0.199 for Cep~A~(W).
\label{FIG:Larson}}
\end{figure*}

  Figure~\ref{FIG:Larson} shows the deviations of the line center
velocity (velocity dispersion), averaged over sizes chosen from the
wavelet spectrum, as a function of the size of each region.  Such
deviations of the line center velocity, averaged over sizes chosen
from a wavelet spectrum, have been previously used by \citet{Gill96}
to study turbulence in molecular clouds.  In this figure we can
appreciate a large dispersion of values.  However, points tend to
clump together around certain regions of the diagrams.

  A closer examination of Fig.~\ref{FIG:Larson} for Cep~A~(E) shows a
cluster of points at position (20,9) corresponding to structure
labeled G (see Fig.~\ref{FIG:Size}), while the points corresponding to
structure C lie just below of them.  Taken separately, each one of
these groups seem to define a line in the diagram, offset vertically
from one another, but of comparable slopes.  The slope of the lines,
in fact, is also similar to the slope of the line that would be
obtained by fitting a line to all the data, $m$=0.26 (indicated with a
solid line in Fig.~\ref{FIG:Larson}).  The dispersion is large, but
there is at least another factor contributing to the amount of
velocity dispersion that has already been mentioned: the distance from
the components of the structure to the outflow source. We take out
this effect by subtracting this systematic linear contribution with
the distance. Thus, in Fig.~\ref{FIG:Larson}, points belonging to the
C group, which are closer to the source, would merge together with
those points of the farther G group.  The dispersion would be lower
and the slope would have a similar value. This defines a relation
similar to Larson's law for molecular clouds \citep{Larson81}.  Larson
found that molecular clouds show a suprathermal velocity dispersion
that correlates with the size $a$ of the cloud as $\sigma_v \propto
a^{0.38}$. Although different values of the power index, $\alpha$,
have been mentioned in the literature, ranging from 0.33, for pure
Kolmogorov turbulence, to 0.5, a more appropriate value in terms of
the Virial Theorem \citep{Goodman98}. In the Cep~A~(E) case we obtain
a slope $\alpha$ of 0.25 which is closer to the Kolmogorov value.

  In the western outflow it is more difficult to identify clumps of
points, and we were not able to find a relation of velocity dispersion
with position either. A least square fit of a line to the data
gives a slope of 0.20.  However, there appears to be two disperse
clumps of points, a lower one and an upper one, each defining a line
with a larger slope $\alpha\sim$0.5 and parallel to each other.  This
latter value would be closer to a case where Virial Theorem applies.

\begin{figure*}
\begin{center}
\includegraphics[width=\textwidth]{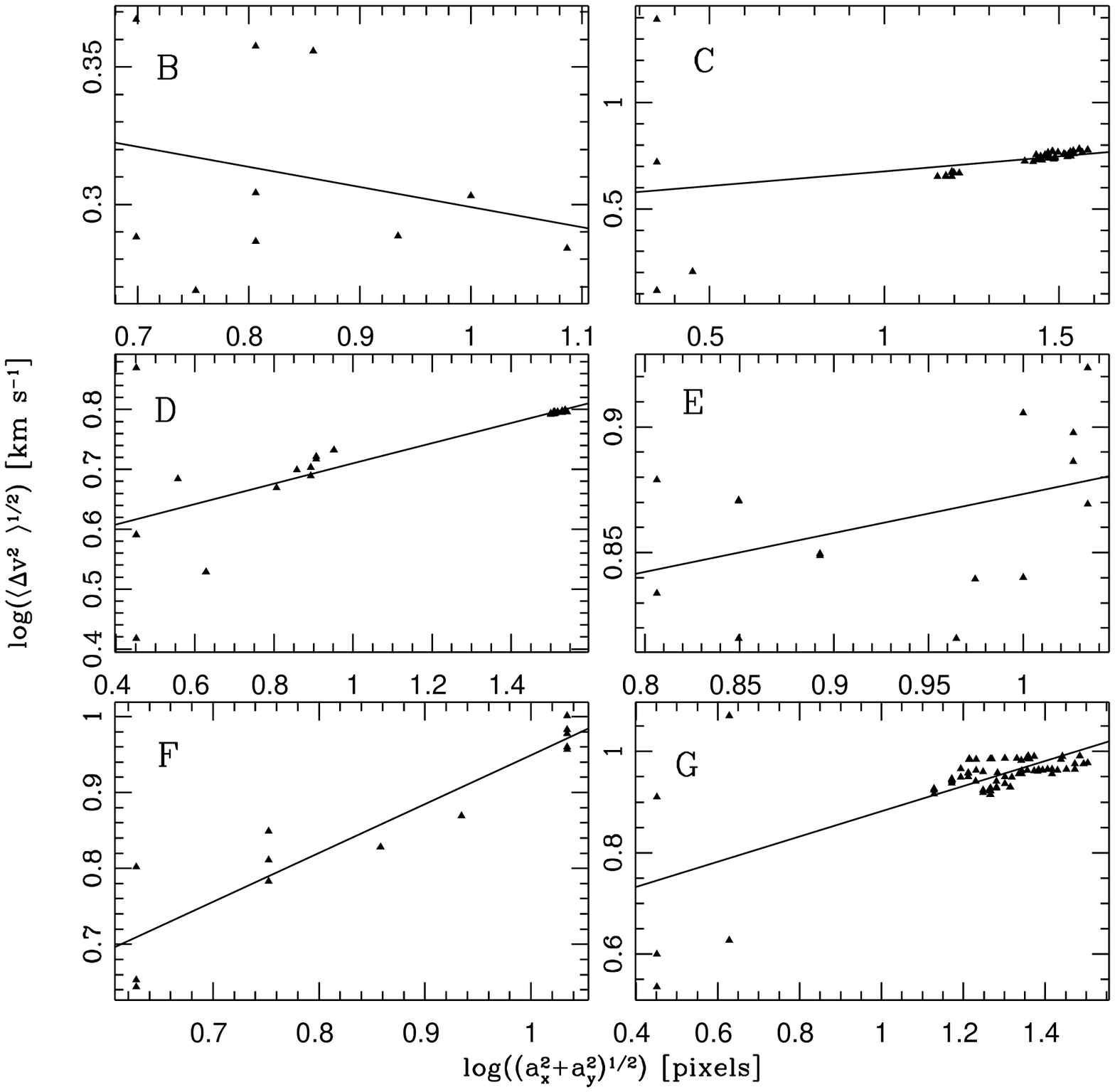}
\end {center}
\caption{ Relation of the deviations of the line center velocity,
 averaged over regions chosen from the wavelet spectrum, $ \langle
 \Delta v^2 \rangle ^{1/2}$, and the size of the region $a_x^2
 +a_y^2)^{1/2}$ for regions {\bf B} to {\bf G} of Cep~A~(E).  The
 solid line is the best fit to the points. The values of the slope
 $\alpha$ of the fits are given in the last column of
 Table~\ref{TAB1}.
\label{FIG:Elarson}}
\end{figure*}

\begin{figure*}
\begin{center}
\includegraphics[width=\textwidth]{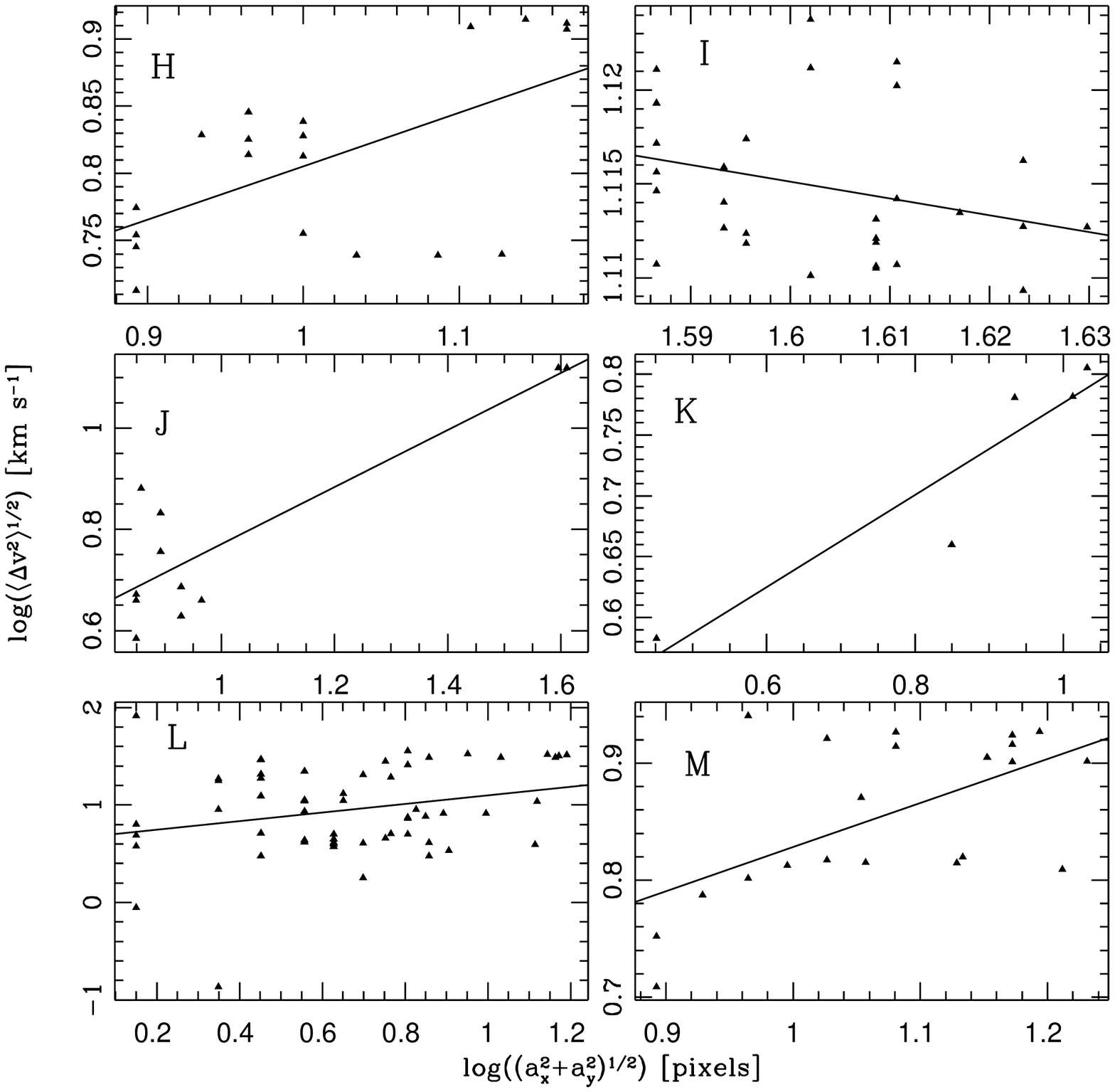}
\end {center}
\caption{ Relation of the deviations of the line center velocity,
 averaged over regions chosen from the wavelet spectrum, $ \langle
 \Delta v^2 \rangle ^{1/2}$, and the size of the region $(a_x^2
 +a_y^2)^{1/2}$ for regions {\bf H} to {\bf M} of Cep~A~(W).  The
 solid line is the best fits to the points. The values of the slope
 $\alpha$ of the fits are given in the last column of
 Table~\ref{TAB1}.
\label{FIG:Wlarson}}
\end{figure*}

  The analysis of the individual H$_2$ condensations
(c.f. Fig.~\ref{FIG:Channel-map}) in the Cep~A~(E) and Cep~A~(W) is
illustrated in Fig.~\ref{FIG:Elarson} and \ref{FIG:Wlarson}.  We show
the deviations for the line center velocity, averaged over sizes
chosen from the wavelet spectrum for each condensation. For each H$_2$
condensation the values of the best fit slope $\alpha$ are presented
in the last column of Table~\ref{TAB1}. The Cep~A~(E) clumps yield a
mean $\alpha$ of 0.21$\pm$0.21 while the Cep~A~(W) yield a slightly
higher value of 0.34$\pm$0.20. However, some knots show poor
correlations than others (see plots for knots B, E, H and I). Using
only knots C, D, F and G for Cep~A~(E) and knots J, K, L and M for
Cep~A~(W) yields an $\alpha$ of 0.30$\pm$0.20 and 0.44$\pm$0.08, which
support a Kolmogorov case in the first and a more virialized region
for the western case.
%figure 13

%figure 14

%figure 15

\section{Conclusions}
\label{sec:Conclusions}

  We have presented the velocity structure of the molecular hydrogen
outflows from the regions Cep~A~(E) and Cep~A~(W) obtained from the
H$_2$ $v$=1--0 $S$(1) doppler shifted line emission at 2.12~$\mu$m. Both
the velocity channel maps and the integrated H$_2$ image show a
complex structure of 12~individual clumps along two separated
structures oriented roughly in the east-west direction.

  Given the complexity of these structures, we have carried out an
anisotropic wavelet analysis of the H$_2$ image, which automatically
detects the position and characteristic sizes (along and across the
region axis) of the clumps.

\begin{enumerate}

\item There is evidence for a Mach disk in Cep~A~(E). The efflux point
is located at the center of a bow shock structure and we measure
blue-shifted velocities of 22 to 28~km~s$^{-1}$. This observation
implies that a molecular jet is driving the outflow.

\item Cep~A~(W) on the other hand, is consistent with a hot bubble in
expansion driving C-shocks. We presented the kinematic gradient of
one of such shocks as an example.

\item The H$_2$ flux-velocity relation is present in both
outflows. The break velocity of the eastern outflow is lower than that
of the western outflow, and we have argued that this is indicative of
greater turbulence in Cep~A~(E).

\item The wavelet analysis has confirmed and quantified trends
observed in the centroid velocity measurements: that the eastern
outflow shows a constant line center velocity and an increasing
velocity dispersion, while the western outflow shows a velocity
gradient and a constant velocity dispersion. The larger velocity
dispersion and gradient in the eastern outflow is taken as indicative
of turbulence, and allows us to conclude also that turbulence
increases with distance.

\item Suggestive propositions about the kind of turbulence present in
both outflows, are extracted from an analysis of the relation of the
velocity dispersion as a function of the size of the structures
(cells) identified as unities by the wavelet spectrum. Using only
knots with a good correlation yields an $\alpha$ of 0.30$\pm$0.20 for
Cep~A~(E) and 0.44$\pm$0.08 for Cep~A~(W), which support a Kolmogorov
case in the first and a more virialized region for the western case.

\end{enumerate}

\acknowledgements

We thank to L. F. Rodr\'{\i}guez for his valuable comments.  Our
appreciation to the Observatorio Astron\'omico Nacional (OAN/SPM)
staff for their assistance and technical support during the
observations, most specially the night assistants G. Garc\'{\i}a and
F. Montalvo.  I. Cruz-Gonz\'alez and L. Salas acknowledge support from
CONACyT research grant 36574-E. We thank the referee, Antonio
Chrysostomou, for his useful comments to improve this paper.

\clearpage

\end{document}